\documentclass[preprint,showpacs,nofootinbib,amsmath,amssymb,tightenlines]{revtex4}
\usepackage{amsmath, amssymb, amsbsy, amsfonts, amsthm}
\usepackage[mathscr]{eucal}
\usepackage{bm}

\theoremstyle{plain}
\newtheorem{theorem}{Theorem}
\newtheorem{proposition}[theorem]{Proposition}
\newtheorem{lemma}[theorem]{Lemma}

\theoremstyle{definition}

\theoremstyle{remark}

\theoremstyle{plain}
\newcommand{\startproof}{\setlength{\parindent}{0in}\textbf{Proof.} }
\newcommand{\finishproof}{\hfill $\blacksquare$ \\}

\newcommand{\cfier}{C}

\newcommand{\C}{\mathbb{C}}

\newcommand{\R}{\mathbb{R}}

\newcommand{\dif}{\mathrm{d}}
\newcommand{\deriv}[2]{\frac{\dif #1}{\dif #2}}
\newcommand{\pderiv}[2]{\frac{\partial #1}{\partial #2}}
\newcommand{\half}{\frac{1}{2}}
\newcommand{\nt}[1]{\mathring{#1}} % puts an 'o' above the argument symbol

\newcommand{\U}{\mathrm{U}}
\newcommand{\SO}{\mathrm{SO}}
\newcommand{\SU}{\mathrm{SU}}
\newcommand{\SL}{\mathrm{SL}}
\newcommand{\Diff}{\mathrm{Diff}}

\newcommand{\mIm}{\mathrm{Im}\,} % For both image and imaginary part, depending on the context.

\newcommand{\su}{\mathfrak{su}}
\newcommand{\so}{\mathfrak{so}}
\newcommand{\msl}{\mathfrak{sl}} %`sl' was already defined as a command

\newcommand{\forget}{\mathfrak{F}\,}
\newcommand{\Hil}{\mathcal{H}}
\newcommand{\Cyl}{\mathrm{Cyl}}
\newcommand{\Dom}{\mathrm{Dom}\,}

\newcommand{\mspan}{\mathrm{span}}
\newcommand{\perpin}[1]{(\perp \; in \; #1)}

\newcommand{\dummy}{\rule{0mm}{0mm}}

\newcommand{\scrA}{\mathcal{A}}
\newcommand{\scrD}{\mathcal{D}}

\newcommand{\scrC}{\mathcal{C}}

\newcommand{\scrU}{\mathcal{U}}

\newcommand{\scrE}{\mathscr{E}}

\newcommand{\lcylstar}{\big(}
\newcommand{\lcylstarS}{\lcylstar}
\newcommand{\lcylstarft}{\big(} % for footnotes
\newcommand{\rcyl}{\big\rangle}
\newcommand{\rcylS}{\rcyl}
\newcommand{\cylmid}{\big|}
\newcommand{\cylmidft}{\big|}  % for footnotes

\newcommand{\lstarbr}{\Big[}
\newcommand{\rstarbr}{\Big]}

\newcommand{\finann}{\alpha}
\newcommand{\lqgz}{\mathfrak{Z}}
\begin{document}

%\title{b-embeddings and the robustness of the basic quantization in LQC}
%\title{On the physical interpretation of states in loop quantum cosmology}
\title{Relating loop quantum cosmology to loop quantum gravity: Symmetric sectors and embeddings}
\author{J. Engle}
\email{engle@cpt.univ-mrs.fr}
\affiliation{
Centre de Physique Th\'{e}orique de Luminy
\footnote{Unit\'{e} mixte de recherche (UMR 6207) du CNRS et des
Universit\'{e}s de Provence (Aix-Marseille I), de la Mediterran\'{e}e
(Aix-Marseille II) et du Sud (Toulon-Var); laboratoire affili\'{e} \`{a}
la FRUMAM (FR 2291).}\\
CNRS, F-13288 Marseille EU
}
%\date{}

\begin{abstract}
In this paper we address the meaning of states in
loop quantum cosmology (LQC), in the context of loop
quantum gravity. First, we introduce a rigorous formulation
of an embedding proposed by Bojowald and Kastrup, of LQC
states into loop quantum gravity. Then, using certain holomorphic
representations, a new class of embeddings,
called b-embeddings, are constructed, following the ideas
of \cite{engle}. We exhibit a class of operators
preserving each of these embeddings, and show their
consistency with the LQC quantization. In the b-embedding case,
the classical analogues of these operators separate points in
phase space.  Embedding at the gauge and diffeomorphism
invariant level is discussed briefly in the conclusions.
\end{abstract}

\pacs{03.70.+k, 04.60.Ds, 04.60.Pp}
% These numbers refer, respectively, to
% ``Theory of quantized fields'', ``Canonical quantization'',
% and ``Loop quantum gravity, quantum geometry, spinfoams''

\maketitle

\section{Introduction}

An important recent development in physics is the construction of a
model of quantum cosmology concretely related to a background
independent approach to full quantum gravity. This theory of
cosmology goes under the name of loop quantum cosmology (LQC), and
consists in a quantization of the cosmological sector of general
relativity, using variables and quantization techniques analogous to
those used in loop quantum gravity (LQG) -- a concrete,
background independent approach to full quantum gravity.

However, beyond similarity of methods of quantization, it is not a
priori clear that LQC accurately reflects the cosmological sector of
LQG (the meaning of which is even not fully clear).
To answer this question, a first step is to
propose a \textit{definition} of the ``symmetric sector'' of LQG.
One can then ask: can LQC states be embedded into this symmetric sector?
Let us state this question more concretely.
%
% An important question in any theory is the physical interpretation
% of states.  If one does not know the physical interpretation of
% states, one does not know what is being claimed in the theory.
%
The `position variable' in LQG is an $\SU(2)$ connection $A^i_a$.
Symmetric connections are those of the form $A^i_a = c\, \nt{A}^i_a$
(see \S \ref{lqc_rev} of this paper).  Thus, states of LQC are wave
functions of a single real variable $c \in \R$, whereas states of
LQG are wave functions of a full $\SU(2)$ connection $A^i_a$.  One
can then ask: given a wave function $\psi(c)$ in LQC, what is the
wave function $\Psi(A^i_a)$ which it represents? There are different
approaches to this question depending on the approach one takes to
the meaning of `symmetric state' in quantum gravity.
In the foundational work \cite{bksymm}, it is proposed that a state
$\Psi(A^i_a)$ be considered `symmetric' if it is \textit{zero} every
where except on symmetric connections $A^i_a = c\nt{A}^i_a$. This
leads to an obvious strategy for embedding LQC states $\psi(c)$ into
LQG states $\Psi(A^i_a)$, reviewed in \S
\ref{sect_bkembed},\ref{sect_bkrig} of this paper. We refer to this
approach to symmetry as `c'-symmetry, where `c' stands for
`configuration' and refers to the fact that symmetry is imposed only
on the configuration field (i.e. `position field') and not on the
momentum field
(see \cite{engle})\footnote{
In the paper \cite{engle}, this approach to symmetry was
referred to as `A' symmetry. We now refer to this approach to
symmetry as `c'-symmetry, in order to give it a more meaningful
name.}.
In \cite{engle}, a second approach to imposing symmetry was proposed
in which symmetric states have more `spread' and symmetry is imposed
on \textit{both} the configuration and momentum fields in a balanced
way. This approach to symmetry was referred to as `b'-symmetry,
where `b' is now taken to mean `balanced'.  As one might guess,
coherent states can be viewed as playing a role in the definition of
`b'-symmetry. Indeed, as we shall see in this paper, given a family
of (complexifier) coherent states, one can, in a very elegant and
clean way, realize a corresponding `b' embedding of LQC states into
LQG.  This is the main result of the paper. Each `b'-embedding
allows a large class of new operators (the classical analogues of
which separate points in phase space) to be directly carried over to
LQC from full LQG. One finds full consistency with the existing
quantization in LQC. Ultimately, however, one needs to define an
embedding of physical LQC states into the space of physical LQG
states, as these are the only states expected to ultimately describe
nature. We will not fully answer this question in this paper, but will
make some comments regarding it in the conclusions.

For deeper conceptual discussions and analysis regarding the meaning of
symmetry in quantum field theory, and more analysis on
associated embeddings of reduced model states into full quantum
field theories, see \cite{engle, englethesis}
(see also \cite{bojowaldlrr})\footnote{
The recent work \cite{koslowski} discusses the relation between
observable algebras in full and reduced quantum theories, though it does not
discuss the issue of the meaning of symmetric state.
}.
For example, it is worthwhile
to recall from \cite{engle, englethesis, bojowaldlrr} why the most obvious definition
of `symmetric state' fails for quantum gravity (in the spatially compact case), so
that alternatives must be considered.  The `obvious' definition of `symmetric state' is that of a
state invariant under the action of a symmetry group.  However, in spatially compact quantum
gravity, all physical states are expected to be invariant under all
diffeomorphisms. Thus \textit{all} physical states are expected to be invariant under
all actions of spatial symmetry groups, whence the `obvious' notion of symmetry is physically
vacuous in this case. One is forced to consider other approaches to the notion of symmetry.
It is argued in \cite{engle, englethesis, bojowaldlrr} that these other notions of symmetry
(such as `c'-symmetry and `b'-symmetry) are in a certain sense more appropriate anyway.

A further remark is in order.  This work does not address deviations
from LQC due to effects of inhomogeneities. It rather addresses the
meaning of the \textit{exactly} homogeneous (isotropic) sector of
LQG and its relation to LQC --- issues which have barely begun to be
addressed.

The paper is organized as follows.  In section \ref{lqglqc_rev} we
review the basics of LQG and LQC needed for this paper; this section
will also serve to fix notation.  We then introduce holomorphic
representations of LQG and LQC in section \ref{sect_holrep}.  In
section \ref{sect_embeds}, after presenting and making precise the
original embedding proposed by Bojowald and Kastrup (the
`c'-embedding), the holomorphic representations are then used to
construct the `b'-embeddings. Preservation of the `c'-embedding and
of each of the `b'-embeddings by operators, and consistency with the
LQC quantization is discussed in section \ref{sect_opspres}.  We
then conclude with a review of the significance of these results,
and discuss some open questions, in particular the issue of
embedding LQC into LQG at the level of physical states.

\section{Brief review of kinematical structure of LQG and LQC}
\label{lqglqc_rev}

% Note: In these presentations, always keep the mention of
% $\overline{\scrA}$ and $\overline{R}_{Bohr}$ secondary, so that
% it is understood from the start that cylindrical functions are
% first of all functions on $\scrA$ and $\R$, and then can be
% extended (uniquely) by continuity to $\overline{\scrA}$ and
% $\overline{\R}_{Bohr}$ if desired.

\subsection{Loop quantum gravity}
\label{lqg_rev}

Loop quantum gravity \cite{alrev, thiemannrev,
rovellibook, smolinrev} is an approach to the quantization of general
relativity which remains background independent, preserving that
insight of general relativity which Einstein, in later reflection,
expressed as the central lesson of general relativity
\cite{einstein1956}.  Mathematically, modern loop quantum gravity
starts from a formulation of general relativity in terms of an
$SU(2)$ connection $A^i_a$ and a densitized triad field
$\tilde{E}^a_i$ conjugate to it \cite{barbero}.
(Here $A^i_a$ more specifically denotes
the components of the $\SU(2)$ connection with respect to the basis
$\tau_i:= -\,\frac{i}{2}\sigma_i$ of the Lie algebra $\su(2)$.)
%
% This is the definition of $\tau_i$ in both alrev and the
% ``Mathematical structure of LQC'' paper, and is such that
% $[\tau_i,\tau_j]=\epsilon_{ijk}\tau_k$.
%
Let $\scrA$ denote the space of smooth $\SU(2)$ connections, and let $\Gamma$
denote the appropriate phase space of smooth pairs $(A^i_a, \tilde{E}^a_i)$.
The Poisson brackets on $\Gamma$ are given by
\begin{equation}
\{A^i_a(x), \tilde{E}^b_j(y)\} = 8 \pi \gamma G \delta^i_j \delta^b_a
\delta^3(x,y)
\end{equation}
where $G$ is Newton's constant, and $\gamma \in \R^+$ is the
Barbero-Immirzi parameter. The strategy of quantization (at the
kinematical level) is essentially characterized by requiring Wilson
loop functionals to have well-defined operator analogues in the
quantum theory \cite{rs_loops}.  More precisely, the algebra of elementary
configuration variables is chosen to consist in
(real analytic\footnote{
One has some freedom in the precise definition of cylindrical functions.
In the present paper, we require real analyticity
to ensure the existence of the complex analytic
continuations used in section \ref{sect_holrep}.
}) functions of finite
numbers of holonomies of the connection $A^i_a$.  Such functions are
called \textit{cylindrical} and the space of such functions is
denoted $\Cyl$.  The elementary momentum variables are taken to be
the \textit{flux} integrals of the triads: given any 2-surface $S$
and any $\su(2)$-valued function $f$ on it, define the corresponding
\textit{flux} by
%
% I decided to use the definition of flux used in Carlo's book
% in order to avoid the use of differential forms.  The reason for
% this choice is the following.
%
% -> It was the only equation in the whole paper using differential forms.
%    Thus, by removing differential forms from this one equation, I've
%    removed all differential forms from the paper easily.
%    This is good because particle physicist do not generally use
%    differential forms, but they do use distributions.
%
% -> Another reason why this particular equation was a bad place to
%    have the only mention of differential forms: it was near the beginning
%    and so would have needlessly scared away people who are not familiar
%    with differential forms.
%
% -> Lastly, Carlo's definition makes it clear why it is called "flux."
%
% The way to see that this definition is coordinate independent is
% to note that on substituting in $n^a$, $\tilde{E}^a_i \epsilon_{abc}$
% has 3-D density weight zero.  I did not include this in the text because
% the purpose of this review is supposed to be a brief reminder of the
% structure of LQG, not to prove this or that quantity is well-defined.
%
% (Just as a reminder, the definition of flux I was using before was
%
% E(S,f) := \int_S f^i \tilde{E}_i \lrcorner \undertilde{\eta}
%
% where $\lrcorner$ denotes contraction of a vector with the first
% index of a differential form, and $\underline{\eta}$ denotes the
% Levi-Civita tensor of density weight -1.)
%
\begin{equation}
\label{fluxdef}
E(S,f) := \int_S f^i \tilde{E}^a_i n_a \dif \sigma_1 \dif \sigma_2
\end{equation}
where $n_a:= \epsilon_{abc} \pderiv{x^b}{\sigma_1} \pderiv{x^c}{\sigma_2}$,
$(\sigma_1, \sigma_2)$ are arbitrary coordinates on $S$, $x^a$ are arbitrary
coordinates on the spatial manifold, and $\epsilon_{abc}$ denotes the fully
anti-symmetric symbol (i.e., the Levi-Civita tensor of density weight -1).
$\Cyl$ together with the fluxes $E(S,f)$,
considered as phase space functions, generate the elementary algebra
of observables.

For the details of quantization, we primarily refer the reader to,
e.g., \cite{alrev, thiemannrev, rovellibook}.  Nevertheless, because
the structure of the kinematical Hilbert space is used repeatedly in
this paper, we briefly review it here. The Hilbert space of
states, $\Hil$, can be expressed as an $L^2$ space over a certain
distributional extension of $\scrA$ referred to as the space of
\textit{generalized connections},
$\overline{\scrA}$ \cite{ai1992,al_repth,al_diffgeom}.
$\overline{\scrA}$ can be characterized \cite{al_repth, al_diffgeom}
as the space of (arbitrarily
discontinuous) maps $A$ from the space of piecewise analytic curves
to $\SU(2)$ satisfying
\begin{equation}
\label{groupoidmorph} A(e_2 \circ e_1) = A(e_2) \circ A(e_1) \quad
{\rm and} \quad A(e^{-1}) = A(e)^{-1} .
\end{equation}
Note every cylindrical function on $\scrA$ naturally extends to
$\overline{\scrA}$ in an obvious fashion.  We shall use a single
symbol `$\Cyl$' to denote the space of cylindrical functions,
whether they are thought of as functions on $\scrA$ or
$\overline{\scrA}$.
%
% The following two sentences are used in the section on the original
% BK embedding.  The Gel'fand description of both \overline{\scrA}
% and \overline{\R}_{Bohr}, and the fact that we use the associated
% topology, is needed in the section on the original BK embedding.
%
In fact, $\overline{\scrA}$ can be characterized
as the Gel'fand spectrum of $\overline{\Cyl}$, the closure of $\Cyl$
in the sup norm \cite{ai1992, al_repth, alrev}.
This characterization gives $\overline{\scrA}$ a
natural topology. With respect to this topology, one can construct a
diffeomorphism and gauge invariant, regular, faithful, Borel measure
$\mu^o$ on $\overline{\scrA}$ called the Ashtekar-Lewandowski
measure (see \cite{ai1992, al_diffgeom}). The kinematical Hilbert
space of states $\Hil$ can then be taken to be
$L^2(\overline{\scrA},\dif \mu^o)$.
On this space of states $\Hil$, the operators corresponding to
functions in $\Cyl$ act by multiplication, and the operators
$\hat{E}(S,f)$ corresponding to fluxes act by certain
derivations \cite{al_area}.
For further details, see, e.g., \cite{alrev, thiemannrev}.

Finally, we note that $\Cyl^*$, the algebraic dual of $\Cyl$, is
often a convenient home for distributional states. We will use the
following notation: elements $\Phi \in \Cyl$, when considered
abstractly as states, will be denoted $\cylmid \Phi \rcyl$, and
elements of $\Cyl^*$ will be denoted in a fashion such as $\lcylstar
\Psi \cylmid$. The evaluation of an element of $\Cyl^*$ on a test
function $\Phi \in \Cyl$ will be denoted by simple juxtaposition:
\begin{equation}
\lcylstar \Psi \cylmid \Phi \rcyl .
\end{equation}
We have
\begin{equation}
\Cyl \hookrightarrow \Hil \hookrightarrow \Cyl^*
\end{equation}
where the second embedding is the standard antilinear one given by
$\Psi \mapsto \langle \Psi, \cdot \, \rangle$. We will follow the
convention that capital letters are used for states in LQG.
(Lowercase letters will be used for states in LQC.)
When it is clearer to think of a given element $\lcylstar \Psi
\cylmid \in \Cyl^*$ as a `generalized function', we will denote the
generalized function by $\Psi(A)$, so that we symbolically write
\begin{equation}
\lcylstar \Psi \cylmid \Phi \rcyl = \int_{\overline{\scrA}} \dif
\mu^o \overline{\Psi} \Phi .
\end{equation}
%
% Note: the above relation is needed to fix the
% way we identify ``generalized functions'' with
% distributions: that is, the Ashtekar Lewandowski
% measure is the measure used in this identification.
%
% Although some readers could probably guess this,
% its better to at least have it clearly implicit
% as in the equation above, just to be fully clear.
%
% For those who prefer to think of generalized functions
% as elements of Cyl^*, this equation then also serves as
% the *definition* of the symbol representing a
% generalized function. So, it is important.
%

\subsection{Isotropic loop quantum cosmology}
\label{lqc_rev}

Isotropic loop quantum cosmology is an attempt to quantize the
cosmological sector of general relativity using the same techniques
used in loop quantum gravity.  In the original papers \cite{lqcorig,
bksymm}, it was presented to be more than this -- to be a theory of
the cosmological sector of loop quantum gravity itself, at the
kinematical level. We will discuss this viewpoint in \S
\ref{sect_embeds}. However, for the purposes of this section, it
will be clearer to think of LQC as a quantization using methods that
are analogous to loop quantum gravity methods.
The modern kinematics of LQC were formulated in \cite{abl2003}.
For an in-depth
%
% I checked on webster.com, and this is in fact spelled with
% a hyphen.
%
review of LQC, see \cite{bojowaldlrr}. For a shorter but more
recent summary of the field, see \cite{ashtekar_cosm}.

We begin by reviewing the homogeneous-isotropic sector
of general relativity in terms of connection variables.
The presentation will differ from \cite{abl2003}, in that we will
be fixing a particular action of the symmetry group and defining
symmetric connections to be those invariant under this action.
This will be essential for the next sections, and is inspired
by the original approach of \cite{bksymm}.
It is sufficient for the modest purposes of
this paper to consider the spatially flat case, so that the spatial
symmetry group is the Euclidean group.  To define the desired
sector, we must first fix a particular action of the Euclidean
group, $\scrE$, on the $\SU(2)$ principal fiber bundle of the
theory. Concretely, this can be done by fixing an action of $\scrE$
on the basic variables through a combination of spatial
diffeomorphisms and local $\SU(2)$ gauge rotations.  In specifying
the action of $\scrE$, let us use its structure $\scrE \cong \R^3
\rtimes \SO(3)$, so that a typical element will be denoted $(x,r)$.
Also, recall that each element $r \in \SO(3)$ is naturally related
to two elements of $\SU(2)$ (via the standard 2 to 1 homomorphism),
related to each other by a sign. Let $\Lambda(r) \in \SU(2)$ be one
of these two elements of $\SU(2)$ (the sign will end up not
mattering)\footnote{
One could also replace the Euclidean group $\R^3 \rtimes \SO(3)$
with its universal cover $\R^3 \rtimes \SU(2)$; this would allow
the sign ambiguity in $\Lambda(\cdot)$ to be avoided. However, as mentioned,
the sign ambiguity does not matter here.
%
% Note: one knows that $\R^3 \rtimes \SU(2)$ is the universal
% cover of $\R^3 \rtimes \SO(3)$ because (1) it is a cover,
% (2) it is connected and simply connected, and (3) the map from
% $\R^3 \rtimes \SU(2)$ to  $\R^3 \rtimes \SO(3)$ making the former
% a cover is a homomorphism. (The first two conditions are what
% defines something to be a topological universal
% cover.  They are sufficient to uniquely determine the cover
% upto homeomorphism, and thus one can speak of ``the''
% universal cover of a given manifold.  The third condition means
% that the group structure of  $\R^3 \rtimes \SU(2)$ is the
% unique one making it the universal covering group.
% see the wikipedia pages ``covering space'' and ``covering group''.
%
}.
Then define $\scrE$'s action through the following
combination of diffeomorphisms and gauge transformations:
\begin{equation}
\label{euc_action}(x,r) \mapsto \Lambda(r) \circ \phi_{(x,r)}
\end{equation}
where $\Lambda(r)$ here denotes the action of the constant gauge
transformation taking everywhere the value $\Lambda(r) \in \SU(2)$,
and $(x,r) \mapsto \phi_{(x,r)} \in \Diff(M)$ denotes some fixed
action of the Euclidean group on the spatial manifold $M$. The
action induced on the basic variables $(A^i_a, \tilde{E}^a_i)$ is
then
\begin{equation}
(x,r)\cdot(A^i_a, \tilde{E}^a_i) = \left(r^i{}_j
\left(\phi_{(x,r)}\right)_* A^j_a \,\, , \,\, (r^{-1})^j{}_i
\left(\phi_{(x,r)}\right)_* \tilde{E}^a_j \right)
\end{equation}
where $r^i{}_j$ denotes the adjoint action of $r \in \SO(3)$ on
$\so(3) \cong \su(2)$.

Let $\scrA_S$ and $\Gamma_S$ denote the subspaces of $\scrA$ and
$\Gamma$, respectively, consisting in elements invariant under this
action.  $\scrA_S$ is then one dimensional and $\Gamma_S$ is two
dimensional.  For convenience, fix an $\nt{A}^i_a$ in $\scrA_S$ for
reference, and fix a triad $\nt{e}^a_i$ such that
\begin{equation}
\label{bktriaddef} \nt{e}^a_i \nt{A}^j_a = \ell_o^{-1} \delta^j_i.
\end{equation}
for some $\ell_o$ with dimensions of length. Let $\nt{q}_{ab}$
denote the metric determined by $\nt{e}^a_i$. Then
$\left(\nt{A}^i_a, \sqrt{\nt{q}} \nt{e}^a_i\right)$ is in
$\Gamma_S$.  Any other pair $(A^i_a, \tilde{E}^a_i) \in \Gamma_S$
takes the form
\begin{equation}
\label{symmform}
A^i_a = c \nt{A}^i_a, \; \tilde{E}^a_i = p \ell_o^{-2} \sqrt{\nt{q}}
\nt{e}^a_i .
\end{equation}
Here the density weight of $\tilde{E}$ has been absorbed into the
determinant of $\nt{q}_{ab}$.  The factor of $\ell_o$ is inserted in
order to match the conventions dominant in the literature
\cite{abl2003}.  (More precisely, to match the conventions, e.g., in
\cite{abl2003}, set $V_o = \ell_o^3$ and $\nt{\omega}^i_a = \ell_o
\nt{A}^i_a = \nt{e}^i_a$.) Via the coordinates $c$ and $p$, $\scrA_S
\cong \R$ and $\Gamma_S \cong \R^2$.

Following \cite{abl2003}, we take the Poisson brackets to
be\footnote{There is an ambiguity in the definition of the symplectic
structure of the reduced theory, due to the fact that one cannot
integrate the symplectic current over all of $M$, but must restrict
integration to an arbitrary fixed region $\mathcal{V}$.  Following
\cite{abl2003}, this ambiguity has been absorbed into the
definitions of $c$ and $p$, and is in fact the origin of the need
for the parameter $\ell_o$ or $V_o$.}
\begin{equation}
\{c,p\} = \frac{8\pi \gamma G}{3}
\end{equation}
where $\gamma$ denotes the Barbero-Immirzi parameter. As one can
check, every pair $(A^i_a, \tilde{E}^a_i)$ of the form
(\ref{symmform}) automatically satisfies the Gauss and
diffeomorphism constraints. Furthermore, in the form
(\ref{symmform}), the Gauss and diffeomorphism gauge freedoms have
been completely gauge-fixed; this was implicitly accomplished when
we chose one action of the many possible actions of the
Euclidean group.

Let us next define the basic variables for the quantization.
As in the case of loop quantum
gravity, the `loop'-like character of the quantization is encoded in
the choice of a basic configuration algebra based on holonomies.
More precisely, we take the reduced configuration algebra to be the
same as the LQG configuration algebra (evaluated on $\Gamma_S$),
except that we furthermore restrict the edges to be
\textit{straight} with respect to $\nt{q}$.  This simplifies
analysis, as holonomies of connections in $\scrA_S$ along straight
edges take a particularly simple form. Given an oriented straight
edge $e$, let $\hat{e}^a:= \dot{e}^a / |\dot{e}|$, and let $\mu_{e}$
denote the length of $e$ with respect to $\nt{q}$. One then has
\begin{equation}
A(e) = h_{\hat{e}}(e^{\frac{i}{2}\mu_{e} c})
\end{equation}
where $h_{\hat{e}}:\U(1) \rightarrow \SU(2)$ is defined by
\begin{equation}
\label{hdef} h_{\hat{e}}\left(e^{\frac{i \mu c}{2}}\right):= \cos
\frac{\mu c}{2} + 2 \sin \frac{\mu c}{2} (\hat{e} \cdot
\nt{A}^i \tau_i) .
\end{equation}
Thus such holonomies depend on the connection $c$ only via
exponentials $e^{\frac{i\mu c}{2}}$.
Next, we choose to define the space of cylindrical functions
in the reduced theory as follows.
First, let
us define $\underline{\Cyl}$ to be the space of cylindrical
functions in the full theory based on graphs with only straight
edges.  Let $r$ denote the embedding $\R \hookrightarrow \scrA$
given by $c \mapsto c \, \nt{A}^i_a$. Then define
$\Cyl_S := r^*\left[ \, \underline{\Cyl} \, \right]$.
We call $\Cyl_S$ the space of \textit{LQC cylindrical
functions}.
%
% The LQC papers define $\Cyl_S$ as the space generated
% by sums of products of (\ref{}); this is strictly speaking
% a smaller space than what I have defined here.  But
% what I have defined here results in the nice characterization
% (\ref{}) exactly analogous to the characterization in the full
% theory.  Further, when my definition of \Cyl_S is completed
% in the sup norm, or when the usual definition is completed in
% the sup norm, the result is the almost periodic functions.
% (That Cyl_S closed in the sup norm gives the a.p. fns is noted
% in a footnote below.)
% Some LQC papers also say that they take the space of a.p.
% functions as Cyl_S, showing there is variation in the literature
% (or looseness).
%
The elements $\phi: \R \rightarrow \C$ of $\Cyl_S$ are each of the
form
\begin{equation}
\phi(c) = F\left(e^{\frac{i\mu_1 c}{2}}, \dots, e^{\frac{i\mu_n
c}{2}}\right)
\end{equation}
for some $F: \U(1)^n \rightarrow \C$ real analytic and
$\mu_1, \dots, \mu_n \in \R$.
The real numbers $\mu_1, \dots, \mu_n$ play the role
analogous to edges in full LQG, and $e^{\frac{i\mu_1 c}{2}}, \dots, e^{\frac{i\mu_n c}{2}}$
may be viewed as playing the role analogous to the holonomies.
Because of this, we will sometimes imitate the notation used in
full LQG, and write $c(\mu):= e^{\frac{i \mu c}{2}}$ for all $\mu \in \R$.

The momentum variables are obtained by simple restriction to
$\Gamma_S$ of the momentum variables in the full theory.
Substituting (\ref{symmform}) into (\ref{fluxdef}) leads to
\begin{eqnarray}
\nonumber E(S,f) &=& p \int_S f^i \nt{\Tilde{E}}^a_i n_a \dif
\sigma_1 \dif \sigma_2
\\
&=& p \nt{E}(S,f)
\end{eqnarray}
so that the momentum algebra in the reduced theory consists in
multiples of $p$.

For the quantization of these variables, we primarily refer the
reader to \cite{abl2003}. Nevertheless,
because of its importance for this paper,
we briefly review here the structure of the kinematical Hilbert space.
The kinematical Hilbert space $\Hil_S$ can be taken to be
an $L^2$ space over a certain extension $\overline{\R}_{Bohr}$ of the
classical configuration space $\scrA_S \cong \R$. Here
$\overline{\R}_{Bohr}$ is the \textit{Bohr compactification of the
real line}. As with $\overline{\scrA}$, $\overline{\R}_{Bohr}$ can
be arrived at by way of Gel'fand spectral theory. Similar to the full
theory, $\Cyl_S$ consists of bounded functions, so that one can
complete it in the sup norm to obtain a space $\overline{\Cyl}_S$
with the structure of a
$C^*$-algebra\footnote{This
is the algebra of \textit{almost
periodic functions}.}.
$\overline{\R}_{Bohr}$ is
then the Gel'fand spectrum of $\overline{\Cyl}_S$. Furthermore, in
analogy with the characterization of $\overline{\scrA}$ using
(\ref{groupoidmorph}), $\overline{\R}_{Bohr}$ can be identified with
the space of all arbitrarily discontinuous homomorphisms from $\R$
to $\U(1)$, $c: \mu \mapsto c(\mu)$\footnote{This
analogy between the quantum configuration
spaces of LQC and LQG seems to be not explicitly mentioned in prior literature.
%
% (However, see the recent paper \cite{} which came out after the first version
% of this paper.) <-- I decided not to make a big deal about this -- it doesn't
% really matter, and so I decided to omit it.
%
It can be easily seen from the definition of the Bohr compactification given in
\cite{rudin1962}.  The Bohr compactification of an arbitrary locally
compact Abelian group $G$, as defined in \cite{rudin1962}, is just
the space of arbitrarily discontinuous homomorphisms from the dual
group $\Gamma$ to $\U(1)$. In the case $G=(\R,+)$, the dual group is
again $(\R,+)$, whence one has the above characterization of
$\overline{\R}_{Bohr}$. }.
Here $c(\mu)$ generalizes ``$e^{\frac{ic\mu}{2}}$''.  The
construction of $\overline{\R}_{Bohr}$ as the Gel'fand spectrum of
$\overline{\Cyl_S}$ endows $\overline{\R}_{Bohr}$ with a natural
topology. With this topology, the space of continuous functions on
$\overline{\R}_{Bohr}$ is in one to one correspondence with
$\overline{\Cyl}_S$, and $\overline{\R}_{Bohr}$ is compact. Lastly,
$\overline{\R}_{Bohr}$ has the structure of an (Abelian) group; it
thus possesses a unique Haar measure, which we will denote by
$\mu^o_{Bohr}$. $\Hil_S$ can then be taken to be
$L^2(\overline{\R}_{Bohr}, \dif \mu^o_{Bohr})$.
On this Hilbert space, the operators corresponding to cylindrical
functions $\phi$ act by multiplication, and $\hat{p}$
acts as $-i\left(\frac{8\pi\gamma G}{3}\right)\deriv{}{c}$ on
the dense subspace $\Cyl_S \subset \Hil_S$.

Again, distributional
states naturally live in the algebraic dual $\Cyl_S^*$ of $\Cyl_S$.
In a manner similar to the full theory, we will denote elements
$\phi$ of $\Cyl_S$ by $\cylmid \phi \rcylS$, and denote elements of
$\Cyl_S^*$ in the fashion $\lcylstarS \psi \cylmid$, except that,
note, in the reduced theory, lowercase letters will always be used
for the labels. Evaluation of a distribution on a test function will
again be denoted by juxtaposition $\lcylstarS \psi \cylmid \phi
\rcylS$. If $\lcylstarS \psi \cylmid$ is an element of $\Cyl_S^*$, we
let $\psi(c)$ denote the associated generalized function on
$\overline{\R}_{Bohr}$, so that symbolically we write
\begin{equation}
\lcylstarS \psi \cylmid \phi \rcylS
= \int_{\overline{\R}_{Bohr}} \dif \mu^o_{Bohr} \overline{\psi} \phi .
\end{equation}

Lastly, returning to the full theory, in this paper
we will have need of the algebraic dual $\underline{\Cyl}^*$ of
$\underline{\Cyl}$, defined above.
Elements of $\underline{\Cyl}^*$
will again be denoted in the fashion $\lcylstar \Psi \cylmid$, and
evaluation on an element $\cylmid \Phi \rcyl \in \underline{\Cyl}$
again by juxtaposition, $\lcylstar \Psi \cylmid \Phi \rcyl$.
What is the meaning of elements of $(\underline{\Cyl})^*$, as compared
to elements of $\Cyl^*$?  Elements of $(\underline{\Cyl})^*$ are just like
elements of $\Cyl^*$ except that they have \textit{less information}: they do
not know how to be evaluated on test functions not in $\underline{\Cyl}$, and that
is the only difference.  One can introduce a `forgetful mapping'
$\forget: \Cyl^* \rightarrow (\underline{\Cyl})^*$ by
\begin{equation}
\label{forgetdef}
\forget \lstarbr \lcylstar \Psi \cylmid \rstarbr \cylmid \Phi \rcyl := \lcylstar \Psi \cylmid \Phi \rcyl
\end{equation}
for all $\cylmid \Phi \rcyl$.  That is, it maps
a given $\lcylstar \Psi \cylmid \in \Cyl^*$ to
the corresponding element in $(\underline{\Cyl})^*$
which simply `forgets' how to be evaluated on test
functions not in $\underline{\Cyl}$. $\forget$ will
be used sometimes this paper.
%
% I decided to omit the following, because it doesn't really clarify anything, but probably
% confuses things.  It doesn't seem (to me, at the present moment) to be a fruitful view point.
% Nevertheless, its definitely worth keeping in mind, and worth keeping as a %-note.
%
% As $\forget$ is onto (but not injective), it provides yet
% another way to think about $(\underline{\Cyl})^*$:
% \begin{equation}
% (\underline{\Cyl})^* = \frac{\Cyl^*}{\Ker \forget}.
% \end{equation}
% Thus, elements of $(\underline{\Cyl})^*$ may also be thought of as
% equivalence classes of elements of $\Cyl^*$.

\section{Holomorphic representations of LQG and LQC}
\label{sect_holrep}

\subsection{Holomorphic representations via coherent state transform}

In this section we review how the complexifier coherent states
introduced in \cite{GCSI_II, thiemann2006} are directly related to certain
holomorphic representations of loop quantum gravity. These
holomorphic representations were first introduced in
\cite{thiemann1996}.
By `holomorphic representation' we mean a
representation of a quantum system in which states are
represented as holomorphic functions. We will introduce
such representations
via a map $U$ from a given `position' (`Schr\"{o}dinger'-like) representation
to the desired holomorphic representation. The innerproduct and
action of basic operators in the holomorphic representation
are then taken to be defined via this map $U$, so that the two
representations are unitarily equivalent by construction.

Let us first introduce the complexifier coherent states, and then the
holomorphic representations to which they lead.

Complexifier coherent states are a generalization of the notion of
coherent state based on the property of each being a simultaneous
eigenstate of all annihilation operators.  The definition of the
coherent states are then in essence determined by the definition of
the annihilation operators.  The annihilation operators, in turn,
are designed to correspond to a complete set of complex coordinates
on phase space.  These complex coordinates are generated from the
real configuration coordinates by means of what is called a
\textit{complexifier}.  To introduce these ideas more concretely,
let us consider a finite dimensional phase space $\Gamma$ taking the
form of a cotangent bundle $T^* \scrC$, and suppose $\{q_1, \dots,
q_N\}$ are a complete set of coordinates on the configuration space
$\scrC$. At the classical level, the complexifier is a positive
function $\cfier: \Gamma \rightarrow \R^+$. The corresponding
complex coordinates on phase space are defined by
\begin{equation}
\label{zdef}
 \finann_i := \sum_{n=0}^{\infty} \frac{i^n}{n!}
\{q_i,\cfier\}_{(n)}
\end{equation}
where $\{q_i,\cfier\}_{(n)}:= \{\dots\{q_i, \cfier\}, \dots
,\cfier\}$, where $\cfier$ appears $n$ times on the right hand side.
By making an additional
assumption on the dependence of $\cfier$ on the momenta, one
obtains, generically, that these complex coordinates separate points
on $\Gamma$ \cite{thiemann2006}. Quantizing (\ref{zdef}) gives
\begin{equation}
\hat{\finann}_i = \sum_{n=0}^{\infty} \frac{1}{n!} [\hat{q}_i,
\hat{\cfier}]_{(n)}
\end{equation}
where $[\hat{q}_i,\hat{\cfier}]_{(n)} :=
[\dots[\hat{q}_i,\hat{\cfier}], \dots ,\hat{\cfier}]$, $n$ times. Using the
Baker-Campbell-Hausdorff formula,
\begin{equation}
\hat{\finann}_i = e^{-\hat{\cfier}}\hat{q}_i e^{\hat{\cfier}}.
\end{equation}
Following \cite{thiemann2006}, we refer to these operators as
\textit{annihilation operators}. We will find it convenient to also
introduce the conjugate coordinates $z_i:= \overline{\finann_i}$,
with quantization
\begin{equation}
\label{zquantiz}
\hat{z}_i:=\hat{\finann}_i^\dagger = e^{\hat{C}} \hat{q}_i e^{-\hat{C}}.
\end{equation}
%
% The fact that this quantization also follows more directly from
% the complex conjugate of the definition of \alpha_i, and
% again using the BCH formula, is too obvious, immediate, and
% irrelevant to mention.
%
Indeed, it
is the analogues of these $z_i$ coordinates which will be most used
in this paper.

For each phase space point $\xi \in \Gamma$, define the corresponding
coherent state to be
\begin{equation}
\label{cohstatedef} \psi^\cfier_{\vec{z}[\xi]}(\vec{q}) :=
\left(e^{-\hat{\cfier}}\delta_{\vec{q}\,'}(\vec{q})\right)_{\vec{q}\,'
\rightarrow \overline{\vec{z}[\xi]}}
\end{equation}
where we have chosen to label the coherent state with the
coordinates $\vec{z}$ for reasons that will become clear shortly.
Here $\delta_{\vec{q}\,'}(\vec{q})$ denotes the Dirac delta
distribution with respect to the measure defining the innerproduct
on the state space, and $\vec{q}\,' \rightarrow \vec{\finann}$
denotes complex-analytic continuation. This state is a simultaneous
eigenstate of the annihilation operators $\hat{\finann}_i$ with
eigenvalues $\finann_i[\xi]=\overline{z_i[\xi]}$.

From these coherent states, one can construct a holomorphic
representation. Let $\scrU \subseteq \C^N$ denote the range of the $N$-tuple
$(\finann_i)$ of complex coordinates.
For each $\Phi \in \Hil$, define the function
$U_{\cfier} \Phi: \scrU \rightarrow \C$ by
\begin{equation}
\label{holrep1} (U_\cfier \Phi)(\vec{z}):= \langle
\psi^\cfier_{\vec{z}}, \Phi \rangle .
\end{equation}
$(U_\cfier \Phi)(\vec{z})$ is then holomorphic in $\vec{z}$.
Substituting equation (\ref{cohstatedef}) into (\ref{holrep1}), and
taking the complex-analytic continuation out of the inner product,
one obtains
\begin{eqnarray}
\nonumber U_\cfier(\Phi) &=& \langle e^{-\hat{\cfier}} \delta_{\vec{q}'},
\Phi \rangle_{\vec{q}' \rightarrow \vec{z}} =
\left(e^{-\hat{\cfier}}\delta_{\vec{q}'}\right)(\Phi)_{\vec{q}' \rightarrow \vec{z}}
\\
\nonumber &:=& \delta_{\vec{q}'}\left(e^{-\hat{\cfier}} \Phi
\right)_{\vec{q}' \rightarrow \vec{z}}
\\
\label{holrep2} &=& \left(e^{-\hat{\cfier}}
\Phi\right)(\vec{q}')_{\vec{q}' \rightarrow \vec{z}} .
\end{eqnarray}
Where, in the first line, we begin by re-interpreting
$e^{-\hat{\cfier}}\delta_{\vec{q}}$ as an appropriate distribution (e.g.,
as a Schwarz distribution, if we assume $\Phi$ is Schwarz). The last
line shows that $U_\cfier$ is the same as the generalized coherent state
transform defined in \cite{thiemann1996}. In fact, historically,
this coherent state transform came first, and the complexifier
coherent states were a later development from the transform.  That
is, historically, the logic of discovery was the opposite of that
presented above. However, the above presentation makes clearer the
physical meaning of the complex coordinates $\{z_i\}$ in terms of
the complexifier.

The operators
$\hat{z}_i$ act by multiplication in this representation.
%
% The argument of the wavefunction is the coordinate cooresponding
% to the *`creation operators'* even in the Bargmann representation.
%
More generally, for all holomorphic functions $F(\vec{z})$ with suitably
bounded growth, we can define $F(\hat{\vec{z}})$ using a power series
expansion of $F$.  Then it is not hard to see that $F(\hat{\vec{z}})$
also acts via multiplication in this representation.
From this property, and the expression (\ref{holrep2}), one obtains a
direct relation between
$F(\hat{\vec{z}})$ and $F(\hat{\vec{q}})$ (the operator corresponding
to the restriction of $F( \cdot )$ to $\R^N \subset \C^N$, evaluated
on $\hat{\vec{q}}$). For,
\begin{eqnarray}
\nonumber
U_\cfier(F(\hat{\vec{z}}) \Phi)(\vec{z}) &=&
F(\vec{z}) (U_\cfier \Phi)(\vec{z}) \\
\nonumber
U_\cfier(F(\hat{\vec{z}}) \Phi)(\vec{q}) &=&
F(\vec{q}) (U_\cfier \Phi)(\vec{q}) \\
(e^{-\hat{C}}F(\hat{\vec{z}}) \Phi)(\vec{q}) &=&
F(\vec{q}) (e^{-\hat{C}}\Phi)(\vec{q})
\end{eqnarray}
whence $e^{-\hat{C}} F(\hat{\vec{z}}) = F(\hat{\vec{q}}) e^{-\hat{C}}$,
that is
\begin{equation}
\label{opdirect}
F(\hat{\vec{z}}) = e^{\hat{C}} F(\hat{\vec{q}}) e^{-\hat{C}}.
\end{equation}
This expression also follows more directly from the
definition of $F(\hat{\vec{z}})$ via a power series expansion,
and (\ref{zquantiz}).
%
% The reason why I present the *above* proof of (\ref{opdirect})
% rather than the one via power series expansion is that
% it is the above definition which will be needed in the
% next subsections.  But I didn't want to mention this
% motivation because I felt that its not sufficiently
% worthy of mention.
%

As a simple example of these constructions,
consider $n$ uncoupled harmonic oscillators. One
then has $\scrC = \R^n$, and $\Gamma = T^* \scrC$.  If we choose
$\half \vec{p}^2$ as our complexifier, the resulting complexifier
coherent states, modulo normalization, are the usual Gaussian
dynamical coherent states for
$n$ uncoupled harmonic oscillators.
%(exercise.)
The corresponding holomorphic representation is related to
the classic Bargmann representation \cite{bargmann} by
\begin{equation}
U_\cfier(\Phi)(\vec{z}) = 2^{-n/2} \pi^{-n/4} \exp\left(-\frac{1}{4}\sum_i
z_i^2\right) U_{Bargmann}(\Phi)\left(\tfrac{1}{\sqrt{2}}\vec{z}\right).
\end{equation}
%(exercise.)

\subsection{Holomorphic representations of LQG}
\label{hollqg_sect}

Let us apply these constructions to kinematical LQG. First, let an
appropriate non-negative function $\cfier$ be chosen as a complexifier,
such that it has a non-negative, self-adjoint operator $\hat{\cfier}$ as
its quantization.  In the literature the spatial volume of the
entire universe is often used, because it is the simplest
spatially-diffeomorphism invariant observable we know. As above,
$\cfier$ is used to construct a complex coordinate $A^\C$ on phase
space:
\begin{equation}
\label{ACdef} \left(A^\C\right)^i_a(x) := \sum_{n=0}^{\infty}
\frac{i^n}{n!} \{A^i_a(x),\cfier\}_{(n)} .
\end{equation}
$A^\C$ is thus a one-form taking values in $\su(2)^\C = \msl(2,\C)$,
whence $A^\C$ may be interpreted as an $\SL(2,\C)$ connection.  We
shall denote the space of $\SL(2,\C)$ connections by $\scrA^\C$, as
it is the complexification of the space $\scrA$ of $\SU(2)$
connections.  For each piece-wise analytic curve $e$, define
the annihilation operator
\begin{equation}
\label{lqgann_def}
\hat{A}^\C(e) := e^{-\hat{\cfier}} \hat{A}(e) e^{\hat{\cfier}} .
\end{equation}
As in the last subsection, we also introduce a conjugate coordinate
$\lqgz$, again an $\SL(2,\C)$ connection, defined by
\begin{equation}
\label{lqgz_def} \lqgz^i_a =
\overline{\left(A^\C\right)^i_a}
\end{equation}
so that
$\left(\lqgz^i_a\tau_i\right)= -\left(\left(A^\C\right)^i_a\tau_i\right)^\dagger$.
Classically this implies\footnote{This
is a non-trivial fact because of the path ordering in
the exponential expression for the holonomy.
In (\ref{lqgz_hol}), essentially the effects of the
$\dagger$ and the inverse on path ordering undo each other, yielding the
stated relation.
}
\begin{equation}
\label{lqgz_hol} \lqgz(e) = \left(A^\C(e)^\dagger \right)^{-1}
\end{equation}
whence one is motivated to define the operator analogues
\begin{equation}
\hat{\lqgz}(e) := \left(\hat{A}^\C(e)^\dagger \right)^{-1} .
\end{equation}
Substituting $(\ref{lqgann_def})$ into this expression, and using
the fact that the eigenvalues of $\hat{A}(e)$ are unitary matrices,
one has
\begin{equation}
\hat{\lqgz}(e)=e^{\hat{\cfier}} \hat{A}(e) e^{-\hat{\cfier}} .
\end{equation}

%
% The logic behind the organization of the paragraph below is
% the following:  The coh. states and the holomorphic representation
% are first introduced as one and the same object.  Then we proceed
% to describe the properties of this object (from its two different
% perspectives).  I think this organization is good, finally!
%

Next, for each phase space point $\xi \in \Gamma$, we again define a
corresponding coherent state:
\begin{equation}
\label{def_lqgcoh}
\Psi^\cfier_{\lqgz[\xi]}(A):=\left(e^{-\hat{\cfier}}\delta_{A'}(A)\right)_{A'\rightarrow
A^\C[\xi]}
\end{equation}
where, similar to the last subsection, we label the coherent states
by the coordinate $\lqgz$. Here $\delta_{A'}(A)$ denotes the Dirac
delta distribution with respect to the Ashtekar-Lewandowski measure
$\mu^o$.  In the present case, these coherent states turn out to be
non-normalizable --- that is, they fail to be elements of $\Hil$.
Therefore,
$\Psi_{\lqgz}^\cfier(A)$ in (\ref{def_lqgcoh}) should be viewed as
a generalized wavefunction. To ensure that this generalized wavefunction
corresponds to an element of $\Cyl^*$, we must demand that $\hat{\cfier}$
preserve $\Cyl$
%
% A footnote mark directly adjacent to a mathematical symbol
% is really confusing, hence the reason for this space.
%
\footnote{This will be the case, for example, if $\hat{\cfier}$ is
pure momentum, as will be stipulated in the next section. }.
The direct expression for $\lcylstar \Psi^\cfier_{\lqgz} \cylmid$
as an element of $\Cyl^*$ then coincides with the expression
for $(U\Phi)(\lqgz)$ defining the holomorphic representation:
\begin{equation}
\lcylstar \Psi^\cfier_{\lqgz} \cylmid \Phi \rcyl
= (U\Phi)(\lqgz) = (e^{-\hat{\cfier}} \Phi)(A')_{A'\rightarrow \lqgz}
\end{equation}
where the final expression is derived in a manner analogous to
(\ref{holrep2}).

The states $\Psi^\cfier_{\lqgz[\xi]}$, as generalized wavefunctions,
are again simultaneous eigenstates of the annihilation operators:
\begin{equation}
\hat{A}^\C(e) \Psi^\cfier_{\lqgz[\xi]} = A^\C(e)[\xi]
\Psi^\cfier_{\lqgz[\xi]}
\end{equation}
for all $e$. As elements of $\Cyl^*$, on the other hand, they are eigenstates
of the dual of the conjugate operators $\hat{\lqgz}(e)$:
\begin{equation}
\hat{\lqgz}(e)^* \lcylstar \Psi^\cfier_{\lqgz}\cylmid \, = \lqgz(e)
\lcylstar \Psi^\cfier_{\lqgz}\cylmid .
\end{equation}

The map $U$ defining the holomorphic representation maps
each $\Phi \in \Cyl$ to a \textit{holomorphic cylindrical function}
--- that is, for each $\Phi \in \Cyl$, $(U\Phi)(\lqgz)$ is of the form
$(U\Phi)(\lqgz) =
\tilde{F}(\lqgz(e_1),\dots,\lqgz(e_n))$ for some $e_1, \dots, e_n$
and some $\tilde{F}: \SL(2,\C)^n \rightarrow \C$
holomorphic.\footnote{As
a side note, in this representation defined by $U$, the
inner product will take a certain form. This then gives us an inner
product on the space of holomorphic cylindrical functions of
$\SL(2,\C)$ connections.  In fact, there is one such inner product
for each choice of complexifier.  If the complexifier is gauge and
diffeomorphism invariant, then this inner product will also be gauge
and diffeomorphism invariant. In self-dual LQG, it is expected one
needs to restrict to holomorphic wavefunctions in order to fix a
natural polarization \cite{ashtekar1991}.  Thus, except for the
reality conditions, these innerproducts are candidate innerproducts
for self-dual LQG \cite{thiemann1996}.  This reminds us that in doing self-dual
LQG, the problem is not so much in finding \textit{a} gauge and
diffeomorphism invariant innerproduct, but in finding one that satisfies
the self-dual reality conditions.  This is a point missed
in some papers over the last several years. }
In this representation, the operators $\hat{\lqgz}(e)$ act by multiplication:
\begin{equation}
\label{lqg_actmult} U\left(\hat{\lqgz}(e)\Phi \right)(\lqgz) =
\lqgz(e)\Big(U\Phi\Big)(\lqgz)
\end{equation}
for all $e$.  It is then natural to define the quantization of
more general
holomorphic functions of $\lqgz$ also by multiplication in this
representation. The most general case in which multiplication by
$F(\lqgz)$ preserves the image of $U$ is when $F(\lqgz)$ is a
holomorphic cylindrical function. Given any such function
$F(\lqgz)$, we can define its quantization
$\widehat{F(\lqgz)}=F(\hat{\lqgz})$ by\footnote{
One can also define $\widehat{F(\lqgz)}=F(\hat{\lqgz})$ via a power series expansion of $F(\lqgz)$ in
matrix elements of holonomies of $\lqgz$. However, this involves more subtleties.
}:
\begin{equation}
U\left(\widehat{F(\lqgz)}\Phi\right)(\lqgz) = F(\lqgz)\Big(U\Phi\Big)(\lqgz) .
\end{equation}
As in (\ref{opdirect}), this definition of $\widehat{F(\lqgz)}$ is equivalent
to the more direct expression
\begin{equation}
\label{lqg_opdir}
\widehat{F(\lqgz)} = e^{\hat{C}} F(\hat{A}) e^{-\hat{C}} .
\end{equation}

\subsection{Holomorphic representations of LQC}

The definition of the holomorphic representation of LQC is similar
to those already presented.
Choose a non-negative function $\cfier_S$ on the reduced phase space,
with operator analogue $\hat{\cfier}_S$, non-negative and
self-adjoint\footnote{
In \cite{lqcmaster}, for example,
$\cfier_S$ is taken to be $\half p^2$,
as this yields well understood
Gaussian coherent states.  However, in \S \ref{sect_hols}, we will
motivate a different choice of $\cfier_S$.}.
$\cfier_S$ can then be used to define a complex phase space coordinate
$c^\C$. For each $\mu \in \R$, we then define the operator analogue
of $c^\C(\mu):= e^{\frac{i\mu c^\C}{2}}$ to be
\begin{equation}
\widehat{c^\C(\mu)}:= e^{-\hat{\cfier}_S} \widehat{c(\mu)}
e^{\hat{\cfier}_S} .
\end{equation}
As before, introduce a conjugate coordinate $z:=\overline{c^\C}$.
Defining $z(\mu):=e^{\frac{i\mu z}{2}}$, classically one has
$z(\mu)=\overline{c^\C(\mu)}^{\,\,-1}$, so that we define
\begin{equation}
\widehat{z(\mu)}:= \left(\widehat{c^\C(\mu)}^\dagger\right)^{-1}.
\end{equation}
For each phase space point $\xi \in \Gamma_S$, in a manner similar
to before, one constructs a coherent state $\psi_{z[\xi]}^{\cfier_S}$.
The coherent states are again non-normalizable.
Again, in analogy to the last subsection, in order for the coherent states to
define elements of $\Cyl^*_S$, we must stipulate that
$e^{-\hat{\cfier}_S}$ preserve $\Cyl_S$, which will be the case if
$\hat{\cfier}_S$ is pure momentum, for example.
The direct expression for $\lcylstarS \psi_{z}^{\cfier_S} \cylmid$
as an element of $\Cyl_S^*$, and, simultaneously, the expression
for the map $U$ defining the holomorphic representation,
is given by
\begin{equation}
(U_S\phi)(z):= \lcylstarS \psi_{z}^{\cfier_S} \cylmid \phi \rcyl
= (e^{-\hat{C}_S}\phi)(c')_{c'\rightarrow z}
\end{equation}
for $\phi \in \Cyl_S$.
Each coherent state $\psi^{\cfier_S}_z$ is a simultaneous
eigenstate of the annihilation operators, and the corresponding element
$\lcylstar \psi_z^{\cfier_S} \cylmid$ of $\Cyl_S^*$ is
an eigenstate of the dual of the conjugate operators $\widehat{z(\cdot)}$,
in analogy with the last subsection.
The map $U_S$ defining the holomorphic representation
maps each $\phi \in \Cyl_S$ to a function $(U_S \phi)(z)$
holomorphic in $z$.
%
% more precisely, $U_S \Phi_S$ is a holomorphic function of finitely
% many $z(\mu_1), \dots, z(\mu_n)$.
%
In this representation, the operator analogue $\widehat{z( \cdot )}$
again acts by multiplication:
\begin{equation}
\label{lqc_multop}
U_S(\widehat{z(\mu)}\phi)(z) = z(\mu) (U_S \phi)(z)
\end{equation}
for all $\mu \in \R$.
The quantization of any holomorphic function $F(z)$ (depending on $z$
via finitely many $z(\mu_1), \dots z(\mu_n)$) also acts by multiplication,
and has the more direct expression
\begin{equation}
\label{lqc_opdir}
F(\hat{z}) = e^{\hat{C_S}} F(\hat{c}) e^{-\hat{C}_S} .
\end{equation}

\section{Symmetric sectors and embeddings of LQC into LQG}
\label{sect_embeds}

\subsection{The proposal of Bojowald and Kastrup}
\label{sect_bkembed}

In the foundational paper \cite{bksymm}, Bojowald and Kastrup
proposed a notion of symmetric state in LQG and
an associated embedding of symmetry reduced models, quantized using
loop methods, into LQG.  We briefly review their proposal here.
However we do so only for the specific case of isotropic loop
quantum cosmology, in order to avoid unnecessary mathematical
abstraction.

First, in the Bojowald-Kastrup approach, one defines that a state $\Psi$ in LQG be
`symmetric' if it has \textit{support only on symmetric connections} ($\scrA_S$).
Such symmetric states are always distributional and so
define a subspace of $\Cyl^*$.  This is the symmetric sector a la
Bojowald and Kastrup.

One then proposes the following embedding of LQC states into this sector.
Begin by letting
$r: \R \rightarrow \scrA$
denote the map
\begin{equation}
r: c \mapsto c \nt{A}^i_a.
\end{equation}
In \cite{bksymm} it is suggested that the map $r$ be extended to a
map $\overline{r}: \overline{\scrA_S}=\overline{\R}_{Bohr}
\rightarrow \overline{\scrA}$ ``by continuity.''
\footnote{$\overline{\R}_{Bohr}$ is actually used in describing LQC only in the
later work \cite{abl2003}. I am here presenting the idea of Bojowald
and Kastrup in \cite{bksymm}, appropriately modified to incorporate
the clarifications/emendations
% I also considered using `corrections'; but they are only corrections
% in the physical interpretation of the derivations.
in \cite{abl2003}.}
Let $\Cyl_S$ denote cylindrical functions on
$\overline{\scrA_S}=\overline{\R}_{Bohr}$ and let $\Cyl$ denote
cylindrical functions on $\overline{\scrA}$.  Then the pull back via
$\overline{r}$ is a map $\overline{r}^*: \Cyl \rightarrow \Cyl_S$.
We can then define $\sigma: \Cyl_S \rightarrow \Cyl^*$ by
\begin{equation}
\label{bk_embed}
\lcylstar \sigma \phi \cylmid \Psi \rcyl := \int_{\overline{\R}_{Bohr}}
\dif \mu^o_{Bohr} \overline{\phi} (\overline{r}^* \Psi)
\end{equation}
where $\mu^o_{Bohr}$ is the Haar measure on $\overline{\R}_{Bohr}$.
It is not hard to check, at a heuristic level, that
the elements in the image of $\sigma$ have support only on $\mIm
\overline{r}$, the space of ``symmetric generalized connections.''
Thus $\sigma$ gives a map from LQC states to the symmetric sector
proposed by Bojowald and Kastrup.

%
% I read from the Mathworld website (paraphrased):
%
% The range of a function and the image of a function mean the
% same thing, namely the space of points reached by the values of
% a function.  Furthermore, of these two terms, *Image* is preferred
% by mathematicians (emphasis Mathworld's, if I remember correctly).
% Perhaps this is because the term `Range' is ambiguous in certain
% contexts.
%
% Also, as a side note: it is *`codomain'* that refers to just a
% convenient set containing the image of a function, that is used
% when defining the function.  The range=image is then a subset of
% of the codomain. (I had been confused on this point and had
% accidentally interchanged the meanings of codomain and range in
% my mind.)
%

The problem with this construction is that, in fact, the extended
map $\overline{r}: \overline{\R}_{Bohr} \rightarrow
\overline{\scrA}$, does not exist
\cite{bf2007}.\footnote{In the original presentation \cite{bksymm}, without incorporating
the corrections in \cite{abl2003}, $\overline{r}$ does not exist either, as
in \cite{bksymm}
the quantum configuration space $\overline{\scrU}$ is not a completion of the
classical one $\scrU = \R$.
%
% This is because the configuration algebra in \cite{bksymm} did not separate
% points in $\scrU = \R$.
% (I made this a %-note because this footnote is just a side note, and there's no need
% to overemphasize bk's error.)
}
Fortunately, for the purposes of this paper, we can side step this
issue by modifying the construction of $\sigma$.  This modification
will be presented in the next subsection.

Another, perhaps limiting, aspect of this embedding is the fact that
the states in its image only satisfy a symmetry condition on the
\textit{configuration} variable $A$.  The embedding in subsection
\ref{sect_bembed} will address this issue.

\subsection{A refinement and precise formulation of the Bojowald-
Kastrup proposal: c-symmetry and the c-embedding}
\label{sect_bkrig}
%
% The c-symmetric sector is a refinement of that proposed by
% Bojowald and Kastrup.
% However, the c-embedding is not really a full refinement: it
% is just a precise formulation. Hence the inclusion of both
% in the title of the subsection.
%

\noindent\textit{The c-symmetric sector}

First, in order to have a clearer motivation,
let us propose a definition of (what will be)
the c-symmetric sector using an \textit{operator equation}.
Define $\mathcal{V}_c \subset \Cyl^*$ --- the `c'-symmetric sector ---
to be the set of all
$\lcylstar \Psi \cylmid \in \Cyl^*$ such that, for all $g \in \scrE$
and piecewise analytic edges $e$,
\begin{equation}
\label{csymmeqn} (g \cdot \hat{A})(e)^* \lcylstar \Psi \cylmid =
\hat{A}(e)^* \lcylstar \Psi \cylmid .
\end{equation}
This is the operator version of the classical equation $g \cdot A =
A, \forall g \in \scrE$, imposing symmetry on the connection $A$
(but \textit{not} on the conjugate momentum). Condition (\ref{csymmeqn})
is actually somewhat stronger than requiring $\lcylstar \Psi \cylmid$
to have support only on $\scrA_S$. Thus $\mathcal{V}_c$ is,
strictly speaking, a proper subset of the symmetric sector
proposed by Bojowald and Kastrup.

%
% $\mathcal{V}_c$ (or, rather, its projection via $\forget$)
% will nevertheless still contain the entirety of the
% image of the embedding to be defined next.

\dummy\\
\noindent\textit{The c-embedding}

In constructing the `c'-embedding, let us begin where the construction
in the last subsection breaks down: that is, when one tries to extend
$r: \R \rightarrow \scrA$ by
continuity.  The first step in solving the problem is to accept
the non-existence of the extension of $r$ and realize that we need
to work at the level of functions on $\R$ and $\scrA$, and not at
the level of their compactifications.  As noted in \S\S
\ref{lqg_rev}, cylindrical functions can be thought of as functions
on either $\scrA$ or $\overline{\scrA}$: the algebraic structure of
the associated space of cylindrical functions is the same.  For the
rest of this paper, we will think of cylindrical functions as
functions on the space of smooth connections, $\scrA$. Likewise, as
noted in \ref{lqc_rev}, elements of $\Cyl_S$ can be thought of as
functions on either $\R$ or $\overline{\R}_{Bohr}$; for the rest of
this paper, we will think of them as functions on $\R$.

%
% I decided to define the embedding $\Cyl_S^* \rightarror \underline{\Cyl}^*$
% right from the start for two separate reasons -- each being sufficient.
% (1.) A similar definition will be most natural
% in the next subsection, and its best to keep presentations parallel,
% for readability.
% (2.) This makes this subsection stronger and the motivations for things
% are clearer, which in turn makes it more readable.
%
As in \S\ref{lqc_rev}, let $\underline{\Cyl}$ denote the
space of cylindrical functions on $\scrA$ depending on graphs
\textit{with only straight edges}.  The image of $\underline{\Cyl}$
under the pull-back map $r^*$ is precisely $\Cyl_S$. This allows us
to define $\iota_c: \Cyl^*_S \rightarrow \underline{\Cyl}^*$ by
\begin{equation}
\label{cembed_def} \iota_c \lstarbr \lcylstarS \psi \cylmid \rstarbr
\cylmid \Phi \rcyl := \lcylstarS \psi \cylmid r^* \Phi \rcyl .
\end{equation}
One can check that this $\iota_c$ is furthermore injective,
and thus an embedding. $\iota_c$ then gives a precise formulation of
the embedding envisioned by Bojowald and Kastrup.
We have defined the embedding on all of
$\Cyl_S^*$ to exhibit generality --- not only do we have an
embedding of $\Cyl_S$ or $\Hil_S$, but we have an embedding of
\textit{all of} $\Cyl_S^*$ into LQG.
Of course, the difficulty is that the non-standard space
$\underline{\Cyl}^*$ is the codomain of $\iota_c$.
The present author does not know how to get around this.
This issue persists even if we restrict the domain of
$\iota_c$ to $\Cyl_S$.  Nevertheless, as will be reviewed in the
conclusions section, elements of $\underline{\Cyl}^*$ still
have a satisfactory physical interpretation.

The image of $\iota_c$ can be stated quite precisely:
it is the space of distributional states in $\underline{\Cyl}^*$
that vanish on non-symmetric connections. That is,
\begin{equation}
\label{imagec}
\mIm \iota_c = \Big\{\; \lcylstar \Psi \cylmid \in
\underline{\Cyl}^* \quad \Big| \quad \Phi \mid_{\scrA_{inv}} = 0
\quad \Rightarrow \quad \lcylstar \Psi \cylmid \Phi \rcyl = 0 \;
\Big\} .
\end{equation}
The proof of this is relegated to appendix \ref{images_app}.
This image (which lies in $\underline{\Cyl}^*$) is directly related
to the symmetric sector $\mathcal{V}_c$ defined above:
\begin{equation}
\label{cimage_inproj}
\mIm \iota_c \subset \forget[\mathcal{V}_c]
\end{equation}
where $\forget$ denotes the `forgetful map' introduced in (\ref{forgetdef}).
We suspect that $\mIm \iota_c$ is not only a subset of $\forget[\mathcal{V}_c]$,
but is also equal to $\forget[\mathcal{V}_c]$; but this has not
been proven. The proof of (\ref{cimage_inproj})
may again be found in appendix \ref{images_app}.

The symmetry condition satisfied by the image indicates in what
sense LQC states are being mapped into symmetric states.  The
heuristic formula we started with, in (\ref{bk_embed}), is perhaps
the simplest and most mathematically natural one.  Further
confidence in this embedding will be given by proposition
\ref{cprop} in section \ref{sect_opspres}. There, it will be found
that holonomies along straight edges preserve the image
$\mIm \iota_c$, and their action on $\mIm \iota_c$ is consistent
with the action of the analogous operators in LQC.

% Just a note, in case relevant for the future:
%
% The symmetry condition satisfied by the image indicates in what
% sense LQC states are being mapped into symmetric states. However,
% the image of the embedding is only part of its specification.  For
% any invertible map $f$ on $\Hil_S$, $\hat{\iota}_c := \iota_c \circ
% f$ would be another embedding with exactly the same image.
%
% If we could intertwine a set of operators with *irreducible* action
% on LQC (E.g., if some full theory operator corresponding to $p$
% could be found that is intertwined by iota_c, in addition to the
% holonomies), then this would be sufficient to uniquely fix the $f$
% ambiguity mentioned above (using at one step Schur's lemma).  Another
% possibility is if $\mIm \iota_c$ possess a natural innerproduct
% independent of this ambiguity.  Then it would be natural to
% require $\mIm$ to be isometric, which would restrict $f$ to be
% unitary, which might also be sufficient to give uniqueness.
% However, because my understanding of uniqueness was so incomplete,
% I decided not to say anything about it in the text.
%

\subsection{b-symmetry and the b-embeddings}
\label{sect_bembed}

\noindent\textit{The basic idea, and the b-symmetric sector}

In the paper \cite{engle}, an alternative approach to symmetry and
an associated paradigm for embedding a
symmetry reduced model into a full quantum field theory were
suggested. The associated notion of symmetric quantum state is
referred to as a `b-symmetric' state. The `b' stands for `balanced,'
and refers to the fact that symmetry is imposed in a more `balanced'
way on both configuration and momentum variables (see \cite{engle}).
At the end of \cite{engle}, an application of this `b-symmetry'
paradigm to loop quantum gravity was partially sketched. We complete
in this section the construction of the (kinematical) `b'
symmetric sector and embedding of LQC into LQG.

The idea is to construct a family of coherent states in LQC and a
family of coherent states in LQG, and use these to define a mapping
from the former into the latter.  Heuristically, we want the
embedding to map each coherent state in LQC into the coherent state
in LQG corresponding to the same phase space point.\footnote{Note
these proposals are natural, not only in light of
\cite{engle}, but also in light of \cite{bt2006}, where for such LQG
coherent states (cut-off to a fixed graph), boundedness of the
inverse triad operator was found, thereby reproducing a feature of LQC.
}
Let $\cfier$ and
$\cfier_S$ denote a choice of complexifier in the full and reduced
theory, and let $\left\{\Psi^\cfier_{\lqgz}\right\}_{\lqgz \in
\scrA^\C}$ and $\left\{\psi^{\cfier_S}_z\right\}_{z\in\C}$ denote
the corresponding families of coherent states.
From this description of the embedding, one would expect the image
of the embedding to be
\begin{equation}
\label{heuristicspan} \mspan\{ \Psi^\cfier_{\lqgz[\xi]} \}_{\xi \in
\Gamma_S} ,
\end{equation}
closed in some appropriate topology.
One might therefore desire to take the above
expression (made appropriately precise) as the definition
of the b-symmetric sector.  However, in order to be consistent in spirit
with the definition of the c-symmetric sector given in (\ref{csymmeqn}),
we choose instead to define the b-symmetric sector via an
operator equation. At this point, in order that the desired
operator equation be well-defined, we must stipulate that
$\cfier$ be at least Euclidean ($\scrE$) invariant
(see footnote \ref{fn_welldefops}).
Define $\mathcal{V}_b$ to be the space of
all $\lcylstar \Psi \cylmid \in \Cyl^*$ such that\footnote{
%
% I decided to include the following equation in terms of the $\underline{\Cyl}^*$
% elements, because it is obvious from the above equation, and so
% gives an intermediate step in the proof of the equation in terms of
% the generalized wavefunction. It also *en passant* points out the subtleties
% of dealing with elements of $\Cyl^*$ or $\underline{\Cyl}^*$ and their generalized
% wavefunctions, and the correct physical interpretation of equations of the former.
% This is very important for those who want to understand, but is unimportant for those
% who work at a heuristic level, and would confuse the text if it were put in the main
% text.
%
In terms of $\lcylstarft \Psi \cylmidft$ as an element of
$\underline{\Cyl}^*$, this equation reads
\begin{equation}
\nonumber (g \cdot \hat{\lqgz})(e)^* \lcylstarft \Psi \cylmidft =
\hat{\lqgz}(e)^* \lcylstarft \Psi \cylmidft, \quad \forall
g\in\scrE.
\end{equation}
}
%
% This space between the footnotes was added because directly
% adjacent footnote marks are confusing.
%
\footnote{\label{fn_welldefops}Here, $(g \cdot \hat{A}^\C)(e)$, and
also $(g \cdot \hat{\lqgz})(e)$, are well-defined as operators for
the following reason.  First, note that the stipulated
$\scrE$-invariance of $\cfier$ implies that $A^\C$ and $\lqgz$ transform
as $\SL(2,\C)$ connections under $\scrE$.  It is not hard to show
this from equations (\ref{ACdef}) and (\ref{lqgz_def}) that give
$(A^\C)^i_a(x)$ and $\lqgz^i_a(x)$ in terms of $A^i_a(x)$ and $\cfier$.
%
% As I say here, this is something easy to verify explicitly, and as this is just a side
% note in a footnote, it would be bad to say more.  But for the record,
% it is easy to prove: one need only start with the expressions for $(A^\C)^i_a(x)$
% and $\lqgz^i_a(x)$ in terms of $A^i_a(x)$ and $\cfier$.
% As elements of $\scrE$ act on phase space functions via pull-back,
% one has
% (g \cdot A^\C)^i_a(x) = (A^\C)^i_a(x)[(g \cdot A), (g \cdot \C)].
% Under the assumption that $\cfier$ is invariant under $\scrE$, $g \cdot \cfier = \cfier$
% for all $g \in \scrE$. For $g \in \scrE$, the above equation for $(g\cdot A^\C)^i_a(x)$
% is then easily seen to reduce to the usual transformation law for a connection.
% In showing this, it is easiest if one replaces the Lie algebra index $i$ with
% two matrix indicies, working in the fundamental representation; for then one can
% use the usual physicists formula for the transformation of a connection (which is simpler)
% and at the same time the way in which one can move the $\Lambda(g)$ matrices out of the
% Poisson brackets freely (b/c they are constant on phase space) is made clearer.
%
% The argument for $(g \cdot \lqgz)^i_a(x)$ is similar.
%
Because $A^\C$ and $\lqgz$ transform as $\SL(2,\C)$ connections
under $\scrE$,
\begin{eqnarray}
\nonumber (g\cdot A^\C)(e) &=& \Lambda(g) A^\C(\phi_g \cdot e)
\Lambda(g)^{-1} \\
\nonumber (g\cdot \lqgz)(e) &=& \Lambda(g) \lqgz(\phi_g \cdot e)
\Lambda(g)^{-1}
\end{eqnarray}
for all $g \in \scrE$. (See (\ref{euc_action}) for notation.) Thus
$(g \cdot A^\C)(e)$ and $(g \cdot \lqgz)(e)$ can be expressed in
terms of the holonomies $A^\C(e)$ and $\lqgz(e)$, which have
well-defined quantizations.
}
\begin{equation}
\label{bsymmeqn} (g \cdot \hat{A}^\C)(e) \Psi = \hat{A}^\C(e) \Psi
\end{equation}
for all $g\in \scrE$ and piece-wise analytic $e$. This is an
operator version of the classical equation $g \cdot A^\C = A^\C,
\forall g \in \scrE$. This classical equation in fact implies the
symmetry of \textit{both} the connection and its conjugate momentum.
This is the sense in which the elements of $\mIm \iota_b$ are
symmetric, and it expresses the `balanced' way in which this
symmetry is imposed --- symmetry is imposed on both configuration
and momenta.  The span in (\ref{heuristicspan}) is a subset of
$\mathcal{V}_b$ so defined. Furthermore, from the definitions of
$\mathcal{V}_c$ and $\mathcal{V}_b$, and from equation
(\ref{lqgann_def}) (and the transformation property noted in
footnote \ref{fn_welldefops}), it is immediate that
\begin{equation}
\label{bc_sectors}
\mathcal{V}_b = e^{-\hat{\cfier}^*}[\mathcal{V}_c] .
\end{equation}
That is, using the terminology of \cite{thiemann2006}, $\mathcal{V}_b$ may be obtained
by simply applying the ``smoothening operator'' $e^{-\hat{\cfier}^*}$ to $\mathcal{V}_c$.

\dummy\\
\noindent\textit{The b-embedding}

To define the embedding, we will
need a map $s:\C\rightarrow\scrA^\C$ defined by
\begin{equation}
s(z[\xi])=\lqgz[\xi]
\end{equation}
for all $\xi \in \Gamma_S \subset \Gamma$. That is, $s$ is just
the representation of the inclusion map $\Gamma_S \hookrightarrow
\Gamma$ in the complex coordinates $z$ and $\lqgz$.
At this point, we stipulate three things: firstly that $\cfier$ and $\cfier_S$
be chosen to be \textit{pure momentum}, secondly that
$\frac{\delta\cfier}{\delta{\tilde{E}^a_i(x)}}$ and $\deriv{\cfier_S}{p}$
vanish only at zero momentum\footnote{
This will be the case if the condition
on growth with respect to momentum in \cite{thiemann2006} is satisfied.
%
% The condition in \cite{thiemann2006} is that the complexifer function
% should grow faster than linearly with respect to any (? was ambiguous
% -- but that means its perfectly okay, as I made the assumption explicitly
% in the text) component of the momentum.
%
}
and lastly that $\cfier$ and $\cfier_S$ be chosen such that $s$ is
\textit{holomorphic}. These assumptions are necessary for the
following ``derivation'' of $\iota_b$.
They are furthermore necessary in proving the intertwining
proposition (\ref{bprop}) in section \ref{sect_opspres}.
The last assumption --- holomorphicity of $s$ --- fixes a
relation between $\cfier$ and $\cfier_S$; this relation is studied
in section \ref{sect_hols}.

Let
\begin{equation}
V := \mspan\Big\{ \lcylstarS \psi^{\cfier_S}_z \cylmid \Big\}_{z \in \C} \subset
\Cyl_S^* .
\end{equation}
The strategy is to first define an embedding
$\tilde{\iota}_b:V \rightarrow \underline{\Cyl}^*$, and then
rewrite it in such a way that it becomes obvious how to extend
it to all of $\Cyl^*_S$.
Define $\tilde{\iota}_b: V \rightarrow
\underline{\Cyl}^*$ by
\begin{equation}
\label{start_eqn} \tilde{\iota}_b \lstarbr \lcylstarS
\psi^{\cfier_S}_{z}\cylmid \rstarbr :=
\forget \lstarbr \lcylstar \Psi^{\cfier}_{s(z)} \cylmid \rstarbr
\end{equation}
%
% Note $\{\psi^{\cfier_S}_z\}$ is, strictly speaking, a linearly independent set:
% no element can be expressed as a \textit{finite} linear combination of the
% others --- limits are necessary before one can do that.  Thus, there is no
% risk of inconsistency in the definition of $\tilde{\iota}_b$ above.
% (The only thing such a consistency check would be needed for would be
% to check if $\tilde{\iota}_b$ is continuous; but I am not worrying about
% such things ... this is just motivation for equation (\ref{bembed_final}).
%
where we have used the forgetful map $\forget$ to obtain a state
in $\underline{\Cyl}^*$. (The reason for defining $\tilde{\iota}_b$ as a
mapping into $\underline{\Cyl}^*$
is in order to ensure the well-definedness of the mapping when the domain is
extended to all of $\Cyl_S^*$ later on.)
Thus, for all $\Phi \in \underline{\Cyl}$,
\begin{eqnarray}
\nonumber \tilde{\iota}_b \lstarbr \lcylstarS \psi^{\cfier_S}_{z}
\cylmid \rstarbr \cylmid \Phi \rcyl
&=& \lcylstar \Psi^{\cfier}_{s(z)} \cylmid \Phi \rcyl
\\
\label{bderiv_eqn2} &=& (U\Phi)(s(z)) .
\end{eqnarray}

Next, define $\pi: \Cyl \rightarrow \Cyl_S$ by $\Phi \mapsto
\left(U_S^{-1}\circ s^* \circ U\right)\Phi$. In order for $\pi$ to
be well-defined, we must define the domain of $\pi$ (`$\Dom \pi$')
and have $s$ such that $\left(s^* \circ U\right)[\Dom \pi] \subseteq
\mIm U_S$. The easiest way to do this is to let $\underline{\Cyl}$
be the domain of $\pi$; then the requirements imposed earlier
on $\cfier$ and $\cfier_S$ imply that $\pi$ is well-defined.
More precisely, the stipulation that $\cfier$ be pure momentum
implies $\hat{\cfier}$ will be graph-preserving and hence preserve
$\underline{\Cyl}$, and this combined with the holomorphicity of
$s$ ensures that $(s^* \circ U)[\Dom \pi]$ is contained in $\mIm U_S$.
%
% Note in the section on the holomorphic representations of LQG,
% we only stipulated that $e^{\hat{\cfier}}$ preserve $\Cyl$.
%
(Whether or not $\pi$ can be defined on all of $\Cyl$, and not just
$\underline{\Cyl}$, is not clear.)

Using $\pi$, (\ref{bderiv_eqn2}) can be rewritten
\begin{eqnarray}
\nonumber \tilde{\iota}_b \lstarbr \lcylstarS \psi^{\cfier_S}_{z}\cylmid
\rstarbr \cylmid \Phi \rcyl
&=& (U_S \pi \Phi)(z)
\\
\label{eqn1} &=& \lcylstarS \psi^{\cfier_S}_{z} \cylmid \pi \Phi \rcyl .
\end{eqnarray}
Using this equation's linearity in $\lcylstarS \psi^{\cfier_S}_{z} \cylmid$,
we have
\begin{equation}
\tilde{\iota}_b \lstarbr \lcylstarS \alpha \cylmid \rstarbr \cylmid
\Phi \rcyl = \lcylstarS \alpha \cylmid \pi \Phi \rcyl
\end{equation}
for all $\lcylstarS \alpha \cylmid \in V$.  This then suggests an obvious extension
to an embedding $\iota_b: \Cyl^*_S \rightarrow \underline{\Cyl}^*$
defined on all of $\Cyl^*_S$:\footnote{The
`overcompleteness' result for the coherent states proven in
\cite{GCSI_II} presumably implies $V$ is dense in $\Cyl_S^*$
with respect to some topology. If so, one could perhaps define
the extension $\iota_b$ of $\tilde{\iota}_b$ just by stipulating
that it be continuous with respect to this topology.
}
\begin{equation}
\label{bdef} \iota_b \lstarbr \lcylstarS \alpha \cylmid \rstarbr
\cylmid \Phi \rcyl = \lcylstarS \alpha \cylmid \pi \Phi \rcyl .
\end{equation}
Note this expression for the `b'-embedding is almost identical to that
of the `c'-embedding in the last subsection: the only difference is that
$r^*$ is replaced with the projector $\pi$, defined using
holomorphic representations.

$\iota_b$ can furthermore be written in terms of $\iota_c$.
%
% This will make the well-definedness and injectivity of
% \iota_b manifest.
%
We start by showing how $s$ and $r$
are related, and then show the relation between $\iota_c$ and $\iota_b$.
First, as $\cfier$ and $\cfier_S$ are pure momentum,
using the Poisson brackets in \S \ref{lqg_rev}-\ref{lqc_rev},
we have
\begin{eqnarray}
\label{lqgz_exp} \lqgz^i_a &=& A^i_a - i k\gamma \frac{\delta C}{\delta
\tilde{E}^a_i(x)} (\tilde{E})
\\
\label{zexp} z &=& c - i\frac{k\gamma}{3} \deriv{C_S}{p}(p) .
\end{eqnarray}
Thus, for $\xi \in \Gamma_S$,
\begin{eqnarray}
\nonumber s(z\mid_{\xi}) &:=& \lqgz\mid_{\xi}
\\
\nonumber &=& A^i_a(x)\mid_{\xi} - i k\gamma \frac{\delta C}{\delta
\tilde{E}^a_i(x)}(\tilde{E})\mid_{\xi}
\\
\label{s_expr} &=& c \nt{A}^i_a(x) - i k\gamma \frac{\delta C}{\delta
\tilde{E}^a_i(x)} (p \nt{\tilde{E}})
\end{eqnarray}
For the case $z = c$ real, $\deriv{C_S}{p} = 0$, so that $p=0$,
whence the second term in (\ref{s_expr}) is zero,
leaving\footnote{Note this equation and the holomorphicity of $s$
imply the explicit expression $s(z) = z \nt{A}$. }
\begin{equation}
\label{s_eq_r}
s(c) = c \nt{A}^i_a = r(c) .
\end{equation}
For $\Phi \in \Cyl$ we then have
\begin{eqnarray}
\nonumber
(s^*(U \Phi))(z)
&=& (U\Phi)(s(z)) = (e^{-\hat{\cfier}} \Phi)(A')_{A' \rightarrow s(z)} \\
\nonumber
&=& (e^{-\hat{\cfier}} \Phi)(s(c'))_{c' \rightarrow z}
= (e^{-\hat{\cfier}} \Phi)(r(c'))_{c' \rightarrow z} \\
\nonumber
&=& ((r^* \circ e^{-\hat{\cfier}})\Phi)(c')_{c' \rightarrow z}
= (e^{-\cfier_S}(e^{\cfier_S} \circ r^* \circ e^{-\hat{\cfier}})\Phi)(c')_{c' \rightarrow z} \\
&=& U_S((e^{\hat{\cfier}_S} \circ r^* \circ e^{-\hat{\cfier}})\Phi)(z)
\end{eqnarray}
where the holomorphicity of $s$ and equation (\ref{s_eq_r})
were used. We thus have
\begin{equation}
\pi := U_S^{-1} \circ s^* \circ U = e^{\cfier_S} \circ r^* \circ
e^{-\hat{\cfier}} .
\end{equation}
So that
\begin{equation}
\label{b_intermsof_c}
\iota_b = e^{-\hat{\cfier}^*}\circ \iota_c \circ e^{\hat{\cfier}_S^*} .
\end{equation}
With this expression, the well definedness of $\iota_b$ as a map
$\Cyl_S^* \rightarrow \underline{\Cyl}^*$ is manifest,
and the injectivity of $\iota_b$ manifestly follows from that
of $\iota_c$.

Let us now discuss the image of the above `b'-embedding.
One expression for the image is precisely
\begin{equation}
\label{bimage}
\mIm \iota_b = \Big\{\;\lcylstar \Psi \cylmid\!\!\in
\underline{\Cyl}^* \quad\Big|\quad
\left(U\Phi\right)\!|_{\lqgz[\Gamma_S]}\!=\!0 \quad \Rightarrow
\quad \lcylstar \Psi \cylmid \Phi \rcyl\!=\!0 \;\Big\}.
\end{equation}
This is seen to be analogous to the expression (\ref{imagec}) for
$\mIm \iota_c$.
The proof of (\ref{bimage}) can be found in appendix \ref{images_app}.
It follows that a given $\lcylstar \Psi \cylmid$ is in the image of $\iota_b$
iff
\begin{equation}
\label{bimage2} \lcylstar \Psi^{\cfier}_{\lqgz[\xi]} \cylmid
\Phi\rcyl = 0 \quad \forall \xi \in \Gamma_S \qquad \Rightarrow
\qquad \lcylstar \Psi \cylmid \Phi \rcyl = 0 .
\end{equation}
This suggests one should be able to write
\begin{equation}
\label{bspan} \mIm \iota_b = \overline{\mspan} \Big\{ \lcylstar
\Psi^\cfier_{\lqgz[\xi]}\cylmid \Big\}_{\xi \in \Gamma_S}
\end{equation}
with closure taken in an appropriate topology
related to the condition (\ref{bimage2}). This gives some
hope that the expectation expressed in equation (\ref{heuristicspan})
%
% Its an `expectation' regarding what $\mIm \iota_b$ will be.
%
might be realized, though nothing precise has been proven at the
moment. Lastly, in analogy with (\ref{bc_sectors}), and as follows
immediately from (\ref{b_intermsof_c}),
\begin{equation}
\mIm \iota_b = e^{-\hat{\cfier}^*}[\mIm \iota_c] .
\end{equation}
Applying $e^{-\hat{\cfier}^*}$ to both sides of (\ref{cimage_inproj}),
and using the fact that $\forget \circ e^{-\hat{\cfier}^*} = e^{-\hat{\cfier}^*} \circ \forget$,
we therefore obtain
\begin{equation}
\label{bimage_inproj}
\mIm \iota_b \subset \forget[\mathcal{V}_b].
\end{equation}
I.e., $\mIm \iota_b$ lies in the projection of $\mathcal{V}_b$
onto $\underline{\Cyl}^*$, as one might hope.

A second approach to defining a `b'-symmetric embedding is to
imitate exactly the expression for the `b-embedding' in equation
(4.86) of \cite{englethesis} (using an integral over the reduced
phase space, and a resolution of unity into coherent states). This
second approach, however, is more involved, and furthermore
leads to an embedding of only $\Cyl_S$.

\dummy\\
\noindent\textit{Note on the relation of this presentation to the structures of the `c' approach}

Finally, we note that the `c'-embedding presented in \S
\ref{sect_bkrig} also could be motivated by a construction similar
to that above.  The only difference in the construction would be
that, instead of initially stipulating that $\psi^{\cfier_S}_z$ map
into $\Psi^\cfier_{s(z)}$, one would stipulate that $\delta_{c'}$
map into $\delta_{r(c')}$.  Furthermore, if (\ref{bspan}) is made more precise,
one could probably write an
equation analogous to (\ref{bspan}) for the c-embedding case, with the
coherent states $\left\{\Psi^\cfier_{s(z)}\right\}_{z\in \C}$
replaced by the delta functions $\left\{\delta_{r(c')}\right\}_{c'
\in \R}$.

\section{Consequences of the holomorphicity of $s$ for the choice of $\cfier$ and $\cfier_S$.}
\label{sect_hols}

In constructing $\iota_b$ in the last section, assumptions were made
regarding the choice of complexifiers $\cfier$ and $\cfier_S$ used.
In particular, it was assumed that $\cfier$ and $\cfier_S$ are both
pure momentum, and that $s$ is holomorphic.  In the present section
we investigate the consequences of these assumptions. For this
section, it will be convenient to define $k:=8\pi G$. We have the
following result:
\begin{lemma}
If $\cfier$ and $\cfier_S$ are both pure momentum, so that $\cfier =
F(\tilde{E})$ and $\cfier_S = G(p)$, and $s$ is holomorphic, then $\cfier_S$
is uniquely determined by $\cfier$ up to a multiple of $p$ and a
(physically irrelevant) constant.  More specifically, $s$ is
holomorphic iff $\exists$ a real, fixed $B^i_a(x)$ such that
\begin{equation}
\label{lemmaeq} \frac{\delta F}{\delta
\tilde{E}^a_i(x)}(p\nt{\tilde{E}}) + B^i_a(x)
= \frac{1}{3}G'(p) \nt{A}^i_a .
\end{equation}
\end{lemma}
{\startproof
From (\ref{s_expr}) we have
\begin{equation}
\label{spec_s} s = c \nt{A}^i_a - i k\gamma \frac{\delta F}{\delta
\tilde{E}^a_i(x)} (p \nt{\tilde{E}}) .
\end{equation}
Let $z=:x+iy$, so that $x=c$ and $y = -\,\frac{k\gamma}{3}G'(p)$. Then
\begin{equation}
\pderiv{}{\overline{z}} := \half\left( \pderiv{}{x} + i\pderiv{}{y}
\right) = \half\left( \pderiv{}{c} - i\frac{3}{k\gamma} (G''(p))^{-1}
\pderiv{}{p}\right) .
\end{equation}
Applying this to equation (\ref{spec_s}), we get
\begin{eqnarray}
2 \pderiv{}{\overline{z}} s(z) =
\nt{A}^i_a(x)-3(G''(p))^{-1}\pderiv{}{p}\left[\frac{\delta F
}{\delta \tilde{E}^a_i(x)}(p\nt{\tilde{E}}) \right] .
\end{eqnarray}
$s$ is thus holomorphic iff
\begin{equation}
\label{secord_diff} \pderiv{}{p}\left[\frac{\delta F}{\delta
\tilde{E}^a_i(x)}(p \nt{\tilde{E}}) \right] = \frac{1}{3}G''(p)
\nt{A}^i_a(x) .
\end{equation}
From this, we can see that $G''(p)$ is uniquely determined by $F$,
so that $G(p)$ is uniquely determined by $F$ up to a term of the
form $B p + C$ \,\, ($B, C \in \R$), as claimed. It is useful to do
the first integration of (\ref{secord_diff}) explicitly.
Integrating both sides with respect to $p$, we get
\begin{equation}
\label{diffeq} \frac{\delta F}{\delta \tilde{E}^a_i(x)}(p
\nt{\tilde{E}}) + B^i_a(x) = \frac{1}{3}G'(p) \nt{A}^i_a(x)
\end{equation}
where $B^i_a(x)$ is some fixed one-form taking values in $\R^3$
(it must be real as the other two terms
are real).
%
% Side note, insufficiently related to the goals of this paper to mention,
% but worthy of a %-comment:
% Using (\ref{lqgz_exp}) and (\ref{zexp}) and the fact that, on
% $\Gamma_S$, $A = c \nt{A}$, equation (\ref{diffeq}) is
% equivalent to the condition that $\exist B$ such that
% $\lqgz = z \nt{A}+iB$.  That is, if $\cfier$ and $\cfier_S$ are pure momentum, $s$
% is holomorphic iff it is of the form $s(z) = z\nt{A}+iB$.
%
\finishproof}

For the case $\cfier =$ spatial volume of the universe, let us
determine what corresponding choices of $\cfier_S$ are allowable, in
the sense of being pure momentum and making $s$ holomorphic. First, we recall from
\cite{thiemann1998} that, for $\cfier = F(\tilde{E}) =$ volume of
the universe,
\begin{equation}
\frac{\delta F}{\delta \tilde{E}^a_i(x)} = \half e^i_a(x)
\end{equation}
where $e^i_a(x)$ is the dynamical co-triad field.  For
$\tilde{E}^a_i = p \ell_o^{-2} \nt{\tilde{E}}^a_i$, let us determine $e^i_a(x)$
in terms of $p$ and $\nt{A}$. First, $\det q =
\det\tilde{E}^a_i = p^3 \ell_o^{-6} \det\nt{\tilde{E}}^a_i =
p^3 \ell_o^{-6} \det\nt{q}$, so
that $e^a_i = (\det q)^{-1/2} \tilde{E}^a_i = \ell_o p^{-1/2} \nt{e}^a_i$.
Thus $e^i_a = \ell_o^{-1} p^{1/2} \nt{e}^i_a$.  From equation
(\ref{bktriaddef}), $\nt{A}^i_a = \ell_o^{-1} \nt{e}^i_a$, so that
\begin{equation}
e^i_a = p^{1/2} \nt{A}^i_a .
\end{equation}
Thus, we have
\begin{equation}
\frac{\delta F}{\delta \tilde{E}^a_i(x)}(p\nt{\tilde{E}}) = \half
p^{1/2} \nt{A}^i_a .
\end{equation}
It follows that, in equation (\ref{lemmaeq}), $B^i_a(x)$ must be of
the form $\frac{B}{3} \nt{A}^i_a(x)$. (\ref{lemmaeq}) then gives us
\begin{equation}
\half p^{1/2} + \frac{B}{3} = \frac{1}{3}G'(p) .
\end{equation}
$G(p)$ must therefore be of the form
\begin{equation}
G(p) = p^{3/2} + B p
\end{equation}
for some $B \in \R$ (plus a possible physically irrelevant
constant).  Note that in LQC, the volume is proportional to $p^{3/2}$
\cite{abl2003}.
Thus, except possibly for the overall coefficient, this result for
$\cfier_S=G(p)$ might have been guessed.
The corresponding complex coordinate $z$ is then
\begin{equation}
z = c - i\frac{k\gamma}{2}\left(p^{1/2}+\frac{2}{3}B\right) .
\end{equation}

\section{Operators preserving the embeddings}
\label{sect_opspres}

Suppose $\mathscr{O}$ is a given function on $\Gamma$, and
$\mathscr{O}_S$ denotes its restriction to $\Gamma_S$.  Suppose
furthermore that each of these possess quantizations,
$\hat{\mathscr{O}}$ in the full theory and $\hat{\mathscr{O}}_S$ the
reduced theory.  Then, given an embedding $\iota$ of the reduced
into the full theory, we would want the following consistency
relation
\begin{equation}
\label{consistency}
\hat{\mathscr{O}} \circ \iota = \iota \circ \hat{\mathscr{O}}_S
\end{equation}
to ideally hold.  For if it holds, then, if $\iota$ is used to
\textit{identify} reduced theory states with full theory states,
then $\hat{\mathscr{O}}_S$ is simply the restriction of
$\hat{\mathscr{O}}$ to the `symmetric sector' $\mIm \iota$. When the
above relation holds, we shall say the embedding $\iota$
`intertwines' the operators $\hat{\mathscr{O}}$ and
$\hat{\mathscr{O}}_S$. If all operators used in the reduced theory
were intertwined in this way, the reduced theory would literally be
simply a description of the symmetric sector $\mIm \iota$ of the
full theory.

We begin with a short result for $\iota_c$: $\iota_c$ intertwines
all functions $F(A)$ of holonomies (of $A^i_a$) along straight
edges.
The restriction to straight edges is needed to ensure
$F(r(\hat{c}))^*$ is well-defined on $\Cyl_S^*$ and $F(\hat{A})^*$
is well-defined on $\underline{\Cyl}^*$.
We will then prove that $\iota_b$ intertwines (the dual action of)
all \textit{holomorphic} functions $F(\lqgz)$ of the holonomies of the
\textit{complex} connection $\lqgz$ along straight edges.
Unlike the
algebra intertwined by
$\iota_c$, this latter algebra of observables separates points on both the
full and reduced phase spaces, which makes it a little more satisfactory, as
it is, in this sense, a ``complete'' set of
observables\footnote{However,
note the work \cite{bojinhom}, where, additionally
using an intermediate cubic lattice model and a certain averaging over
operators, the basic approach of (what we call) the c-embedding
is used to obtain the full basic algebra of LQC from LQG.}.
Whether or not more observables can be intertwined by some $\iota_b$ remains to be seen
\footnote{
Note that if a given operator does not preserve $\mIm \iota_b$, that
merely means the given operator is not ``sharp'' on the symmetric sector.
It is important to understand better what operators should be
``expected'' to preserve the symmetric sector --- this might lead to a
criterion for selecting a b-embedding, or another embedding as preferred.
}.

\begin{proposition}
\label{cprop}
$\iota_c$ intertwines functions of finite numbers of holonomies of
the connection $A^i_a$ along straight edges in the manner (\ref{consistency}).
\end{proposition}
{\startproof
Let $F: \scrA \rightarrow \C$ be any function of the form
$F(A)=\tilde{F}(A(e_1),\dots,A(e_n))$ for some $\tilde{F}:
\SU(2)^n \rightarrow \C$ and $e_1, \dots, e_n$
straight.  Then for any $\lcylstarS \psi \cylmid \in \Cyl_S^*$ and
$\cylmid \Phi \rcylS \in \underline{\Cyl}$, we
have
\begin{eqnarray}
\nonumber \left(F(\hat{A})^* \circ \iota_c\right)\lstarbr \lcylstarS
\psi \cylmid \rstarbr \cylmid \Phi \rcyl &:=& \iota_c \lstarbr
\lcylstarS \psi \cylmid \rstarbr \cylmid F(\hat{A}) \Phi \rcyl
\\
\nonumber &:=& \lcylstarS \psi \cylmid r^*(F(A)\Phi) \rcyl =
\lcylstarS \psi \cylmid F(r(c)) r^*\Phi \rcyl
\\
\nonumber &=& F(r(\hat{c}))^* \lstarbr \lcylstarS \psi \cylmid
\rstarbr \cylmid r^* \Phi \rcyl
\\
&=& \left( \iota_c \circ F(r(\hat{c}))^* \right) \lstarbr
\lcylstarS \psi \cylmid \rstarbr \cylmid \Phi \rcyl
\end{eqnarray}
whence $F(\hat{A})^* \circ \iota_c = \iota_c \circ
F\left(r\left(\hat{c}\right)\right)^*$. \finishproof}

\begin{proposition}
\label{bprop}
$\iota_b$ intertwines the dual action of holomorphic functions of finite
numbers of holonomies of
the complex connection $\lqgz$ along straight edges in the manner (\ref{consistency}).
\end{proposition}
{\startproof
%
% For a dual action on $\Cyl^*$ to exist (as in the 1st line of the proof)
% $F(\hat{\lqgz})$ must preserve $\Cyl$.  As we have assumed $e^{-\hat{\cfier}}$
% preserves $\Cyl$, that means \textit{f must be a cylindrical function for this
% starting expression to even be defined}, and furthermore $F$ must be an element of
% $\underline{\Cyl}$ (or, more accurately, the image of $\underline{\Cyl}$ under $U$,
% i.e., a holomorphic function of a finite number of holonomies along straight edges
% (roughly speaking, but accurately enough for our purposes)) for $F(s(\hat{z}))$
% to be well-defined.
%
Let $F: \scrA^\C \rightarrow \C$ be any function of the form
$F(\lqgz)=\tilde{F}(\lqgz(e_1),\dots,\lqgz(e_n))$ for some
$\tilde{F}: \SL(2,\C)^n \rightarrow \C$ holomorphic and $e_1, \dots,
e_n$ straight.  Then for any $\lcylstar \psi \cylmid \in \Cyl_S^*$ and
$\cylmid \Phi \rcyl \in \underline{\Cyl}$, we have
\begin{eqnarray}
\nonumber
F(\hat{\lqgz})^* \lstarbr \lcylstar \iota_b
\psi \cylmid \rstarbr \cylmid \Phi \rcyl &:=& \iota_b
\lstarbr \lcylstar \psi \cylmid \rstarbr \cylmid F(\hat{\lqgz}) \Phi
\rcyl
\\
\nonumber
&=& \lcylstar \iota_c (e^{\hat{C}_S} \psi) \cylmid
e^{-\hat{C}} F(\hat{\lqgz}) \Phi \rcyl
\\
\nonumber
&=& \lcylstar \iota_c (e^{\hat{C}_S} \psi) \cylmid
F(\hat{A}) e^{-\hat{C}} \Phi \rcyl
\\
\nonumber
&=& \lcylstar \iota_c (F(r(\hat{c}))^\dagger e^{\hat{C}_S} \psi) \cylmid
e^{-\hat{C}} \Phi \rcyl
\\
\nonumber
&=& \lcylstar \iota_c (e^{\hat{C}_S} F(s(\hat{z}))^\dagger \psi) \cylmid
e^{-\hat{C}} \Phi \rcyl
\\
&=& \lcylstar \iota_b (F(s(\hat{z}))^\dagger \psi) \cylmid \Phi \rcyl
\end{eqnarray}
Where (\ref{lqg_opdir}), proposition \ref{cprop} above,
the conjugate of (\ref{lqc_opdir}), equation (\ref{s_eq_r}),
and the holomorphicity of $s$ have been used.
Thus
% This equation here I put in display mode because otherwise it gets
% pushed to the rightmost of the line of text above, and it looks
% not so good.  It looks good this way.
\begin{equation}
F(\hat{\lqgz})^* \circ \iota_b = \iota_b \circ F(s(\hat{z}))^\dagger.
\end{equation}
\finishproof}

\section{Summary and Outlook}

\subsection{The embeddings, their properties and their interpretation}

% \textit{**In this subsection, list all the main points you want
% the reader to carry away (i.e., NOT just a summary)**}

Let us review what has been accomplished. We have made rigorous a
proposal of Bojowald and Kastrup, defining a `c-embedding' of LQC
states into LQG. Secondly, we have shown how to define
`b-embeddings' of LQC states into LQG, motivated by a prescription
using complexifier coherent states. The image of the c-embedding lay
in a `c-symmetric sector' involving a symmetry condition on the
configuration variable, whereas the image of the b-symmetric sector
lay in a `b-symmetric sector' involving a symmetry condition
expressed in terms of annihilation operators. In the process of
proposing a b-embedding, we found need to impose some restrictions
on the complexifiers $\cfier$ and $\cfier_S$ (in the full and
reduced theories) used to construct the embedding. For the case in
which $\cfier$ is the spatial volume of the universe, under these
restrictions, the form of $\cfier_S$ was determined, unique up to
addition of a multiple of $p$. Lastly, we showed that the c and
b-embeddings each \textit{intertwine} a large class of operators. In
other words, a large class of operators were exhibited that preserve
the images of the embeddings and that are consistent with the
existing LQC quantization.

The b-embedding furthermore has certain advantages over the c-embedding.
The b-symmetry condition involves \textit{both} configuration and
momenta; its classical analogue implies exact
symmetry in both configuration and momenta. Furthermore,
the class of operators intertwined by the b-embedding
is `complete'
in the sense that the classical analogues of the operators
separate points in phase space.  Finally,
the `b' approach seems to be necessary to complete the second
strategy for handling the Hamiltonian or Master constraint discussed
in the next subsection.

An unusual fact about the embeddings is that they
have $\underline{\Cyl}^*$ as their codomain.  This is not
desireable first and foremost because background structure is
involved in the very definition of $\underline{\Cyl}^*$ as a space.
(In particular, this causes problems below in trying to embed
into diffeomorphism invariant states.)  Nevertheless,
elements of $\underline{\Cyl}^*$ have a physical interpretation.
It is in fact similar to the interpretation of elements of
$\Cyl^*$ as investigated in \cite{shadows}: Given an element
$\lcylstar \Psi \cylmid$ and any graph $\gamma$ with
straight edges, one can \textit{project} $\lcylstar \Psi \cylmid$
onto $\gamma$ to obtain a \textit{shadow}, $\Psi_\gamma$, which
is in fact normalizable and so can be used to compute,
e.g., expectation values. See \cite{shadows} for further discussion
on such `shadow states'. The difference between elements of
$\underline{\Cyl}^*$ and elements of $\Cyl^*$ is then that the former
only know how to cast shadows on graphs with straight edges.
That is, elements of $\underline{\Cyl}^*$ have \textit{less information}.
This observation was already stated in a different way at the end of
\S \ref{lqc_rev} and motivated the introduction of the
`forgetful map' $\forget$.
%
% Decided to cut this, because I'm not sure if its the correct
% interpretation, and its not necessary:
%
% That is, the states of LQC
% do not represent entire states of LQG, but only represent part of
% the information in a state of LQG.
%
%%%%
%
% This is the perfect place to mention this discussion of
% \underline{\Cy}^*: physical interpretation
% has just been brought up, right after mentioning that the
% embeddings are into Cyl^*.  And this reference is just enough.
%

\subsection{Future directions}

Three future directions are in particular worthy of mention.
Each (except for the last) applies to both $\iota_c$ and $\iota_b$;
however, for brevity, we will often speak in terms of $\iota_b$.
The directions are the following.
\begin{enumerate}
\item One can seek other operators preserving
the image of $\iota_b$, beyond holomorphic functions of holonomies of
$\lqgz$.

\item
\label{physstates_item}
One can try to define symmetric sectors of \textit{physical} states
in LQG, and associated embeddings of LQC states into physical LQG states.
The importance of this, and
a general strategy for achieving it, are discussed below.

\item For handling the Hamiltonian (or Master \cite{master}) constraint,
there is
an alternative to direction (\ref{physstates_item}.) above
which may be more feasible.

\end{enumerate}
Each of these is briefly discussed below.

\dummy \\
\textit{Seeking other operators preserving the images}

The first issue is essentially the following.  In section \ref{sect_opspres},
we have verified that $\mIm \iota_b$ is preserved by, and $\iota_b$
in fact \textit{intertwines} the most obvious set of operators one
might try (a similar statement holds for $\iota_c$).
It remains to investigate further operators of more
direct interest, for example: the volume operator, Hamiltonian
constraint (smeared with constant lapse), or Master constraint.
%
% (prob. don't include this; it does not add much, this is not the
% direction I suggest we go anyway, and the point has already been made
% much more clearly in an earlier section anyway:)
%
% "For, this would give us a way to define the quantization of these
% operators on $\Hil_S$ in a way directly descending from, and
% faithfully representing, the full theory."
%
(Note preservation of $\mIm \iota_b$ by the last two of these operators
is relevant only if one chooses to handle the Hamiltonian/Master
constraint according to the third future direction below rather than
the second.)
The freedom in the choice of complexifier $\cfier$ may help to achieve
preservation (or approximate preservation) of $\mIm \iota_b$ by these operators, though this is not yet
clear. Of course, one also needs to understand better when one should
\textit{expect} a given operator to preserve the ``symmetric sector.''

\dummy\\
\textit{Defining symmetric sectors of physical LQG states and associated embeddings}

The second issue is that of defining symmetric sectors of \textit{physical} LQG
states and associated embeddings of LQC states into \textit{physical} LQG states.
The importance of this lies
in the fact that (1.) it is only physical states
that are proposals for describing reality and
(2.) as the full theory Hamiltonian and Master constraints
are well-defined only on gauge and diffeomorphism invariant states,
if one is to compare dynamics in LQC and LQG as in the third future direction
below, it is necessary to first embed into (at least) gauge and diffeomorphism invariant
states.
For a discussion of the various constraints in LQG, including the Hamiltonian constraint, see
\cite{thiemann1998,alrev,thiemannrev,rovellibook},
and for the Master constraint, see \cite{master}.
In the heuristic program in \cite{engle}, it is suggested that
a symmetric sector solving the constraints, and the associated embedding,
be constructed in steps --- that is, one adds one constraint at a time.
\cite{engle} suggests to achieve this by using
group averaging techniques \cite{almmt1995}.  For example,
to obtain the Gauss-gauge invariant `c' and `b' symmetric sectors
$\mathcal{V}^G_c$ and $\mathcal{V}^G_b$,
one simply applies to the kinematical
sectors $\mathcal{V}_c$ and $\mathcal{V}_b$ the group averaging
map for the Gauss constraint. To
obtain the gauge and diffeomorphism invariant symmetric sectors,
one then in turn group averages $\mathcal{V}^G_c$ and $\mathcal{V}^G_b$
with respect to the diffeomorphism constraint.  This is then to
be repeated for the Master constraint (note one needs to use the
Master constraint rather than the Hamiltonian constraint for the last
step if the group averaging prescription is to be used).
Next, let us turn to the embeddings.
First, as in kinematical LQC the Gauss and
diffeomorphism constraints are already solved, one should be able to
embed kinematical LQC states into solutions of the Gauss and
diffeomorphism constraints in LQG.
%
% "again" in the next sentence is used in the sense
% "analogous to what was done before".
%
This is again to be achieved
by composing the kinematical $\iota_c$ and $\iota_b$ with
the group averaging maps for the Gauss and diffeomorphism constraints.
Second, solutions to the
(Hamiltonian or) Master constraint in LQC should be embeddable into
solutions of the (Hamiltonian or) Master constraint in LQG.
(For the definition of the Hamiltonian and Master constraints in LQC,
see \cite{improved_dyn, abl2003} and \cite{lqcmaster}.)
Regarding this final step, however, it is not clear how it can be
achieved using group averaging alone, or what the general strategy should
be; in any case the third future direction below gives an alternative
to this final step.
%
% This is not put in the text because its distracting --- and for the
% interested reader, its easy to figure it out.
%
% If we let $P_{M_S}$ and $P_M$ denote the
% group averaging maps in the reduced and full theories, given an
% embedding $\iota^{Diff}$ into gauge and diffeomorphism invariant states,
% we would want $\iota^{Phys}$ to satisfy $\iota^{Phys} \circ P_{M_S} = P_M \circ \iota^{Diff}$.
% It is not obvious whether such an $\iota^{Phys}$ will exist, but if it exists,
% it is unique, due to $P_{M_S}$ being onto the physical states in the reduced theory.

In the presciptions above, the applications of group averaging are
heuristic, as one would need to be able to group average states in
$\Cyl^*$ (for the sectors) and states in $\underline{\Cyl}^*$ (for
the embeddings). It is easy to see, however, how these can be made
rigorous for the case of the Gauss constraint. The diffeomorphism
constraint is more difficult.  We do not discuss here the embedding
of solutions to the (Hamiltonian or) Master constraint in LQC into
solutions of the full LQG (Hamiltonian or) Master constraint, as we
do not yet have a general strategy for this and it would probably
require first embedding into diffeomorphism invariant states anyway.

Let us begin by letting $P^{(\Cyl^*)}_G$ and
$P^{(\underline{\Cyl}^*)}_G$ denote the group averaging maps on
$\Cyl^*$ and $\underline{\Cyl}^*$, respectively, for the Gauss
constraint. We will see that these can be made well-defined as maps
$P^{(\Cyl^*)}_G: \Cyl^* \rightarrow (\Cyl^*)_{Gauss-inv.}$ and
$P^{(\underline{\Cyl}^*)}_G: \underline{\Cyl}^* \rightarrow
(\underline{\Cyl}^*)_{Gauss-inv.}$. This is made possible by the
following. First, zero is in the discrete part of the spectra of all
of the Gauss constraints, so that the local-$\SU(2)$-gauge invariant
states live again in the kinematical Hilbert space $\Hil$.  More
precisely, and more importantly, the group averaging map $P_G$ for
the Gauss constraint maps $\Cyl$ back into itself. Using this fact,
we have, for all $\lcylstar \Psi \cylmid \in \Cyl^*$ and $\cylmid
\Phi \rcyl \in \Cyl$,
\begin{eqnarray}
\nonumber P^{(\Cyl^*)}_G \lstarbr \lcylstar \Psi \cylmid \rstarbr
\cylmid \Phi \rcyl &:=& \left(\int_{g \in \mathcal{G}} \scrD g
U_{g^{-1}}^* \lstarbr \lcylstar \Psi \cylmid \rstarbr \right)
\cylmid \Phi \rcyl  \\
\nonumber &=& \int_{g \in \mathcal{G}} \scrD g \lcylstar \Psi
\cylmid U_{g^{-1}} \Phi \rcyl = \int_{g \in \mathcal{G}} \scrD g
\lcylstar \Psi \cylmid U_g \Phi \rcyl \\
\label{groupav_deriv} &=& \lcylstar \Psi \cylmid P_G \Phi \rcyl
\end{eqnarray}
where in the last line $P_G$ denotes
group averaging on $\Cyl$, which is just orthogonal
projection onto the gauge invariant subspace of
$\Cyl$. With this projection $P^{(\Cyl^*)}_G : \Cyl^* \rightarrow (\Cyl^*)_{Gauss-inv}$,
one can define (Gauss) gauge-invariant
`c' and `b' symmetric sectors by
\begin{align}
\mathcal{V}_c^G &:= P^{(\Cyl^*)}_G [\mathcal{V}_c] &
\mathcal{V}_b^G &:= P^{(\Cyl^*)}_G [\mathcal{V}_b].
\end{align}
%
% I needed to keep these equations in display mode to
% keep them in parallel with the equations for the
% embeddings below. If they are either both displayed
% or neither displayed, both displayed is better, I think.
%
Because the action of local $\SU(2)$ gauge
rotations on $\Hil$ preserves $\underline{\Cyl}$, $P_G$ furthermore
preserves $\underline{\Cyl}$.  Using this, one can repeat the
derivation (\ref{groupav_deriv}) to define the group averaging map
$P^{(\underline{\Cyl}^*)}_G: \underline{\Cyl}^* \rightarrow (\underline{\Cyl}^*)_{Gauss-inv}$.
Using this map, one can then define
`c' and `b' embeddings into (Gauss) gauge-invariant states by
\begin{align}
\iota_c^G &:= P^{(\underline{\Cyl}^*)}_G \circ \iota_c & \iota_b^G
&:= P^{(\underline{\Cyl}^*)}_G \circ \iota_b .
\end{align}
The fact that the standard $P_G$ preserves $\underline{\Cyl}$
implies that the forgetful map $\forget$ intertwines the action of
$P^{(\Cyl^*)}_G$ and $P^{(\underline{\Cyl}^*)}_G$ (i.e. $\forget \circ P^{(\Cyl^*)}_G =
P^{(\underline{\Cyl}^*)}_G \circ \forget$). From this, and from
(\ref{cimage_inproj}),(\ref{bimage_inproj}), it follows that
\begin{align}
\mIm \iota_c^G &\subseteq \forget[\mathcal{V}_c^G] & \mIm \iota_b^G
&\subseteq \forget[\mathcal{V}_b^G] .
\end{align}

The above procedure does not immediately work for the
diffeomorphism constraint, because the standard group averaging
map $P_{Diff}$ maps $\Cyl$ \textit{out of itself} into $\Cyl^*$.
Thus, if one were to try to do a formal manipulation similar to
(\ref{groupav_deriv}), the last line would be ill-defined, whence it is not
clear how to define a map
$P^{(\Cyl^*)}_{Diff}: \Cyl^* \rightarrow (\Cyl^*)_{Diff}$.
Thus, at least with the present strategy, it is not clear how to define
the diffeomorphism invariant `c' and `b' symmetric sectors.

If one wishes to define \textit{embeddings} into the diffeomorphism-invariant
states, the situation is worse:
As $\underline{\Cyl}^*$ is
the codomain of $\iota_c$ and $\iota_b$, what is again needed
is a group averaging map from $\underline{\Cyl}^*$ to
the desired gauge-invariant states.
However, because the group of diffeomorphisms does not preserve
the space of straight edges, this group does not preserve
$\underline{\Cyl}$ and hence does not even \textit{act} on
$\underline{\Cyl}^*$.  Thus, for the case
of the diffeomorphism constraint, one cannot even
write down a formal expression for the desired group
averaging map $P^{(\underline{\Cyl}^*)}_{Diff}$, so that, on attempting a
formal manipulation similar to (\ref{groupav_deriv}), one is stopped at
the \textit{first} line. The fact that $\underline{\Cyl}^*$
is the codomain of the c- and b-embeddings is the source of
this particular problem.  In turn, recall $\underline{\Cyl}^*$ was
apparently forced on us as the codomain of the embeddings
because $r^*[\Cyl]$ is not equal to $\Cyl_S$; rather,
it is $r^*[\underline{\Cyl}]$ that is equal to $\Cyl_S$.
That is, the problem can be traced to the fact that only
holonomies along straight edges were included in defining
the configuration algebra of LQC.

The question then arises: might it be necessary to enlarge
the basic configuration algebra of LQC to include holonomies along
\textit{all} edges before one can map reduced theory states into
diffeomorphism invariant states? This would be ideal, but the
quantization of the resulting system is a non-trivial task; in particular
the structure of $r^*[\Cyl]$ appears more complicated.
Another possibility is the following. Again we will speak in
terms of $\iota_b$ even though everything to be said also
applies to $\iota_c$. The embedding $\iota_b$
is constructed using a particular fixed action of the
Euclidean group and a particular choice of reference connection
$\nt{A}$ (the latter determining the former).
Thus, by choosing different actions of the Euclidean group
and different associated reference connections $\nt{A}$, one can
construct different embeddings $\iota^{(\nt{A})}_b$.  Each of these
has as codomain $(\underline{\Cyl}_{\rho[\nt{A}]})^*$, where
$\rho[\nt{A}]$ denotes the action of the Euclidean group determined
by $\nt{A}$, and $\underline{\Cyl}_{\rho[\nt{A}]}$ denotes the space
of cylindrical functions based on graphs with edges straight
with respect to the action $\rho[\nt{A}]$ of the Euclidean group.
The idea is to use \textit{all} of these embeddings rather than
only one of them in constructing $\iota^{Diff}_b$, treating all of
the embeddings on ``equal footing.''  Such an approach would
circumvent the background dependence of the space $\underline{\Cyl}^*$
causing the above problem in the first line in constructing $\iota^{Diff}_b$.
However, we do not yet have a precise strategy for implementing such an idea.

%
% I decided to remove this:
%
% because of the more complicated structure of $r^*[\Cyl]$, and
% the fact that $r^*[\Cyl]\oplus (\C p)$ is probably not closed
% under Poisson brackets and so is not suitable as an algebra of
% basic variables.
% \footnote{
% This was pointed out to the author by T Thiemann.
% }

\dummy\\
\textit{The Hamiltonian or Master constraint}

As mentioned,
the Gauss and diffeomorphism constraints are solved prior to quantization
in LQC; thus kinematical LQC states should be embedded, in at least
some approximate sense, into gauge and diffeomorphism invariant LQG states.
The Hamiltonian (or Master) constraint, on the other hand, is \textit{not} solved prior to
quantization in LQC. This makes two options available.
First, one can try to map solutions of the Hamiltonian
(or Master) constraint in LQC into those of LQG, as suggested above.
The second possibility (which may or may not be easier) is to
try to choose a b-embedding such that the embedding
$\iota^{Diff}_b$ (into diffeomorphism and gauge-invariant states) has its image
at least \textit{approximately} preserved by $\hat{H}$ (or $\hat{M}$) in the
full theory.  This will then give one an approximately closed system
$(\mIm \iota^{Diff}_b, \hat{H})$ (or $(\mIm \iota^{Diff}_b, \hat{M})$) which can then be compared with the
LQC system $(\Hil_S, \hat{H}_S)$ (or $(\Hil_S, \hat{M}_S)$).  It should be noted that in this
second approach, only `b' symmetry has a chance to work.  For,
dynamics always mixes configuration and momenta, and thus `c' symmetry,
as it only involves a constraint on the configuration variables,
hasn't a chance to be even approximately preserved by the dynamical
constraint. This reflects what was found
in the investigations of the Klein-Gordon field in \cite{engle}:
with b-symmetry one had, in fact, exact preservation of the image by the
dynamics --- indeed, in a certain sense, the particular choice of
b-embedding was uniquely determined by this requirement \cite{englethesis}
--- whereas with c-symmetry one was far from any preservation by
dynamics.  Of course, in an interacting system such as gravity, one
can at most hope to have approximate preservation by dynamics, roughly
because symmetric and non-symmetric modes interact, in constrast to the
free case studied in \cite{engle}.

\section*{Acknowledgements}

The author would like to thank Abhay Ashtekar for pointing out the
heuristic embedding (\ref{bk_embed}),
%
% After looking at the email Abhay sent me, I realized this is what
% he did. He pointed out the modification of Martin's original embedding
% which incorporates the modifications of the ABL paper.  The revised
% embedding (\ref{bk_embed}) was written by both Martin and Abhay. The form of
% (\ref{bk_embed}) is also identical to that of Martin's original proposed
% embedding.
%
Martin Bojowald for pointing out that the weaker embedding in
\ref{sect_bkrig} is still sufficient for proposition \ref{cprop},
and Christian Fleischhack for a remark helping to prove proposition
\ref{cimageprop} and for pointing out an important error in a prior
draft. Abhay Ashtekar, Martin Bojowald,
Johannes Brunnemann, Alejandro Perez and Carlo Rovelli also gave
helpful comments on a prior draft.
The referees are also thanked for their careful reading
and helpful suggestions. This work was funded in part by NSF grant
PHY-0090091,
an NSF International Research Fellowship under grant OISE-0601844, and the
Eberly research funds of Penn State. The hospitality of the Centre
de Physique Th\'{e}orique in Marseille, where most of this work was
done, is also gratefully acknowledged.

\appendix

\section{The images}
\label{images_app}

\begin{proposition}
\label{cimageprop} The image of the c-embedding is precisely
\begin{equation}
\label{cim_propeqn}
\mIm \iota_c = \Big\{\;\lcylstar \Psi \cylmid\!\!\in
\underline{\Cyl}^* \quad\Big|\quad \forall \Phi\!\in
\underline{\Cyl}, \quad \Phi|_{\scrA_{inv}}\!=\!0
\quad\Rightarrow\quad \lcylstar \Psi \cylmid \Phi \rcyl\!=\!0
\;\Big\}.
\end{equation}
I.e., it is the space of distributions in $\underline{\Cyl}^*$
vanishing outside of $\scrA_{inv}$.
\end{proposition}
{\startproof
\nopagebreak $\lcylstar \Psi \cylmid \in \mIm \iota_c$ iff there
exists $\lcylstarS \alpha \cylmid \in \Cyl_S$ such that $\lcylstar
\Psi \cylmid = \iota_c \lstarbr \lcylstarS \alpha \cylmid \rstarbr$,
i.e. such that
\begin{equation}
\lcylstar \Psi \cylmid \Phi \rcyl = \iota_c \lstarbr \lcylstar
\alpha \cylmid \rstarbr \cylmid \Phi \rcyl = \lcylstarS \alpha
\cylmid r^* \Phi \rcylS \, \quad \forall \Phi \in \underline{\Cyl} .
\end{equation}
Such an $\alpha$ will exist iff $\lcylstar \Psi \cylmid \Phi \rcyl$
depends on $\Phi$ only via $r^* \Phi$ --- that is, only via $\Phi
|_{\scrA_{inv}}$. Thus, $\alpha$ will exist iff
\begin{equation}
\Phi_1 |_{\scrA_{inv}} = \Phi_2 |_{\scrA_{inv}} \quad \Rightarrow
\quad \lcylstar \Psi \cylmid \Phi_1 \rcyl = \lcylstar \Psi \cylmid
\Phi_2 \rcyl .
\end{equation}
This is equivalent by linearity to
\begin{equation}
\Phi |_{\scrA_{inv}} = 0 \quad \Rightarrow \quad \lcylstar \Psi
\cylmid \Phi \rcyl = 0 ,
\end{equation}
giving the desired condition on the right hand side of
(\ref{cim_propeqn}). \finishproof}

\begin{proposition}
\label{bimageprop} The image of the b-embedding is precisely
\begin{equation}
\label{bim_propeqn}
\mIm \iota_b = \Big\{\;\lcylstar \Psi \cylmid\!\!\in
\underline{\Cyl}^* \quad\Big|\quad \forall \Phi\!\in
\underline{\Cyl}, \quad
\left(U\Phi\right)\!|_{\lqgz[\Gamma_S]}\!=\!0 \quad\Rightarrow\quad
\lcylstar \Psi \cylmid \Phi \rcyl\!=\!0 \;\Big\}.
\end{equation}
\end{proposition}
{\startproof
\nopagebreak $\lcylstar \Psi \cylmid \in \mIm \iota_b$ iff there
exists $\alpha \in \Cyl_S^*$ such that $\lcylstar \Psi \cylmid =
\iota_b \lstarbr \lcylstarS \alpha \cylmid \rstarbr$, i.e. such that
\begin{equation}
\lcylstar \Psi \cylmid \Phi \rcyl = \lcylstarS \alpha \cylmid \pi
\Phi \rcyl, \quad \forall \Phi \in \underline{\Cyl}.
\end{equation}
Such an $\alpha$ will exist iff $\lcylstar \Psi \cylmid \Phi \rcyl$
depends on $\Phi$ only via $\pi \Phi$.  This will be true iff
$\lcylstar \Psi \cylmid \Phi \rcyl$ depends on $\Phi$ only via $s^*
U \Phi$, i.e., only via $(U \Phi) |_{s[\C]} = (U \Phi)
|_{\lqgz[\Gamma_S]}$.  That is, $\alpha$ will exist iff
\begin{equation}
(U \Phi_1) |_{\lqgz[\Gamma_S]} = (U \Phi_2) |_{\lqgz[\Gamma_S]}
\quad \Rightarrow \quad \lcylstar \Psi \cylmid \Phi_1 \rcyl =
\lcylstar \Psi \cylmid \Phi_2 \rcyl .
\end{equation}
I.e., iff
\begin{equation}
(U \Phi) |_{\lqgz[\Gamma_S]}=0 \quad \Rightarrow \quad \lcylstar
\Psi \cylmid \Phi \rcyl = 0 .
\end{equation}
\finishproof}

The next proposition shows the relation of $\mIm \iota_c$
to the c-symmetric sector $\mathcal{V}_c \subset \Cyl^*$ defined in equation (\ref{csymmeqn}).
In order to prepare, we make some notational notes and definitions.
First, if
$(U,\langle \cdot , \cdot \rangle)$ is an inner product space,
and $W$ is a subspace of $V$ which in turn is a subspace of $U$,
let $W^{\perpin{V}}$ denote the orthogonal complement of $W$ in $V$.
%
% I decided its unnecessary and pedantic to include this.
%
% I.e.,
% \begin{equation}
% W^{\perpin{V}} := \{ v \in V \quad | \quad \langle v, w \rangle = 0 \quad \forall w \in W \}.
% \end{equation}
%
Furthermore, we note that, in the following proofs,
``$\oplus$'' will sometimes denote only a vector space (internal)
direct sum and not a Hilbert space direct sum.
Lastly define
\begin{equation}
\label{tilde_cdef}
\tilde{\mathcal{V}}_c :=
\Big\{\; \lcylstar \Psi \cylmid \in
\Cyl^* \quad \Big| \quad \Phi \mid_{\scrA_{inv}} = 0
\quad \Rightarrow \quad \lcylstar \Psi \cylmid \Phi \rcyl = 0 \;
\Big\}
\end{equation}
This is the same as the expression (\ref{cim_propeqn})
for the image of $\iota_c$, except that $\underline{\Cyl}$ is replaced by
$\Cyl$. We are now ready to prove the proposition.

\begin{proposition}
\dummy
\begin{enumerate}
\item $\tilde{\mathcal{V}}_c \subseteq \mathcal{V}_c$
\item $\mIm \iota_c = \forget[\tilde{\mathcal{V}}_c]$
\item $\mIm \iota_c \subseteq \forget[\mathcal{V}_c]$
\end{enumerate}
\end{proposition}
{\startproof

\textit{Proof of (1.)}:
Suppose $\lcylstar \Psi \cylmid \in \tilde{\mathcal{V}}_c$.
Let $\Phi \in \Cyl$, $g\in \scrE$, $e$ piece-wise analytic,
and spinor component indicies $A, B \in \{0,1\}$ be given.
Define $\tilde{\Phi} \in \Cyl$ by
\begin{equation}
\tilde{\Phi}(A):= ((g \cdot A)(e) - A(e))^A{}_B \Phi(A),
\end{equation}
so that $\tilde{\Phi}|_{\scrA_{inv}} = 0$.  As
$\lcylstar \Psi \cylmid \in \tilde{\mathcal{V}}_c$,
\begin{eqnarray}
\nonumber
\lcylstar \Psi \cylmid \tilde{\Phi} \rcyl
&=& \lcylstar \Psi \cylmid ((g \cdot A)(e) - A(e))^A{}_B \Phi \rcyl = 0 \\
\lcylstar \Psi \cylmid (g \cdot A)(e)^A{}_B \Phi \rcyl
&=&  \lcylstar \Psi \cylmid A(e)^A{}_B \Phi \rcyl
\end{eqnarray}
for all $\Phi$, $g$, $e$, and $A, B \in \{0,1\}$, so that
$\lcylstar \Psi \cylmid \in \mathcal{V}_c$.
Thus $\tilde{\mathcal{V}}_c \subseteq \mathcal{V}_c$.\\

\textit{Proof of (2.)}:
Containment in the direction $(\supseteq)$ is immediate from
the definition (\ref{tilde_cdef}) of $\tilde{\mathcal{V}}_c$ and the
expression (\ref{cim_propeqn}) for $\mIm \iota_c$ proven above.
To show containment in the direction $(\subseteq)$,
we begin by defining the following subspaces of $\Cyl$:
\begin{eqnarray}
\Cyl_{\sim} &:=& \{ \Phi \in \Cyl \quad | \quad \Phi|_{\scrA_{inv}} = 0 \} \\
\Cyl_{\sim -} &:=& \Cyl_\sim \cap \underline{\Cyl} \\
\Cyl_{\perp -} &:=& (\Cyl_{\sim -})^{\perpin{\underline{\Cyl}}} \\
\Cyl_{\perp \sim} &:=& (\Cyl_{\sim -})^{\perpin{\Cyl_\sim}}
\end{eqnarray}
One has immediately the equations
%
% motivations for the notations \Cyl_{~-}, \Cyl_{\perp -},
% \Cyl_{\perp ~}, \Cyl_{\perp \perp}:
%
% 1.) meaningful binary way of labeling, which
%     matches the structure of the relations among
%     the spaces.
% 2.) \Cyl_{\perp -} is suggestive of an orthogonal
%     complement in \underline{\Cyl} -- which is
%     exactly what it is: the orthogonal complement
%     of \Cyl_{~-} in \underline{\Cyl}.  A similar
%     statement can be made for \Cyl_{\perp ~}.
%
% 3.) \Cyl_{\perp \perp} is suggestive a space
%     orthogonal to both \underline{\Cyl} and
%     \Cyl_{~}, which is exactly what
%     \Cyl_{\perp \perp} is.
%
\begin{eqnarray}
\underline{\Cyl} &=& \Cyl_{\sim -} \oplus \Cyl_{\perp -} \\
\Cyl_\sim &=& \Cyl_{\sim -} \oplus \Cyl_{\perp \sim} .
\end{eqnarray}
Furthermore, one can see that
\begin{equation}
\underline{\Cyl} \cap \Cyl_{\perp \sim} = \{0\} .
\end{equation}
It follows from this that $\mspan\{\underline{\Cyl} \cup \Cyl_{\perp
\sim}\}$ has the (vector space) direct sum structure\footnote{The
two subspaces are not mutually orthogonal, and so one does not have
a Hilbert space direct sum. }
\begin{equation}
\mspan\{\underline{\Cyl} \cup \Cyl_{\perp \sim}\} = \underline{\Cyl} \oplus \Cyl_{\perp \sim} .
\end{equation}
Finally define
\begin{equation}
\Cyl_{\perp \perp} := (\underline{\Cyl} \oplus \Cyl_{\perp \sim})^{\perpin{\Cyl}} .
\end{equation}
This gives us the decomposition
\begin{eqnarray}
\label{cyl_decomp}
\Cyl &=& \underline{\Cyl} \oplus \Cyl_{\perp \sim} \oplus  \Cyl_{\perp \perp}
\end{eqnarray}
where again the direct sums are vector space direct sums, but not
Hilbert space direct sums. With this decomposition, we are ready to
prove containment in the direction ($\subseteq$).  Let $\lcylstar
\Psi \cylmid \in \mIm \iota_c \subset \underline{\Cyl}^*$ be given.
%
% I decided not to use (\alpha| type notation for $\alpha$ because then I would have
% to define the notation, as up to now I've only defined this notation for elements of
% \Cyl^*, \underline{\Cyl}^*, and \Cyl_S^*.  Its not worth it just for one use and
% makes the text much more noisy for nothing, and elements of $\Cyl_{\perp \perp}^*$
% are not even being used to represent states, so the notation would be confusing
% anyway.  The notation below is immediate and hence requires no introduction, and
% so is, from the viewpoints considered here, the best way to present things.
%
Let $\alpha$ be an arbitrary element of $\Cyl_{\perp \perp}^*$, the algebraic dual of
$\Cyl_{\perp \perp}$.
Using (\ref{cyl_decomp}), define $\lcylstar \tilde{\Psi} \cylmid \in \Cyl^*$
by
\begin{equation}
\lcylstar \tilde{\Psi} \cylmid \Phi \rcyl
= \left\{ \begin{array}
{r@{\quad:\quad}l}
\lcylstar \Psi \cylmid \Phi \rcyl & \Phi \in \underline{\Cyl} \\
0 & \Phi \in \Cyl_{\perp \sim} \\
\alpha(\Phi) & \Phi \in \Cyl_{\perp \perp}
\end{array} \right.
\end{equation}
then extending $\lcylstar \tilde{\Psi} \cylmid$ to
all of $\Cyl$ by linearity.  With $\lcylstar \tilde{\Psi} \cylmid$ thus
defined, one has by construction that $\lcylstar \Psi \cylmid
= \forget\lstarbr \lcylstar \tilde{\Psi} \cylmid \rstarbr$, and one can check that
$\lcylstar \tilde{\Psi} \cylmid \in \tilde{\mathcal{V}}_c$.
Thus $\lcylstar \tilde{\Psi} \cylmid \in \forget[\tilde{\mathcal{V}_c}]$,
proving containment in the the $(\subseteq)$ direction. Therefore
$\mIm \iota_c = \forget[\tilde{\mathcal{V}}_c]$.\\

\textit{Proof of (3.)}: Part (3.) follows immediately from parts
(1.) and (2.) of this proposition.

\finishproof}


\begin{thebibliography}{99}

\bibitem{engle}
Engle J 2006 Quantum field theory and its symmetry reduction
\textit{Class. Quantum Grav.} \textbf{23} 2861-2894

\bibitem{bksymm}
Bojowald M and Kastrup H A 2000 Quantum symmetry reduction for
diffeomorphism invariant theories of connections \textit{Class.
Quantum Grav.} \textbf{17} 3009-3043 \\
Bojowald M and Kastrup H A 2001 Symmetric states in quantum geometry
\texttt{gr-qc/0101061}
% talk at 9th Marcel Grossmann meeting, Rome, July 2-8, 2000

\bibitem{englethesis}
Engle J 2006 ``Black Hole Entropy, Constraints, and Symmetry in
Quantum Gravity'' Ph.D. Thesis, Penn State University \\
(\texttt{http://igpg.gravity.psu.edu/archives/thesis/2006/
englethesis.pdf})

\bibitem{bojowaldlrr}
\S 6.1-6.3 of
Bojowald M 2005 Loop quantum cosmology \textit{Living Rev. Rel.}
\textbf{8} 11 [Online Article]: cited [2 Feb. 2007],
\texttt{http://www.livingreviews.org/lrr-2005-11}

\bibitem{koslowski}
Koslowski T 2006 Reduction of a quantum theory \textit{Preprint:}
\texttt{gr-qc/0612138}

\bibitem{alrev}
Ashtekar A and Lewandowski J 2004 Background independent quantum
gravity: a status report \textit{Class. Quantum Grav.} \textbf{21}
R53-R152

\bibitem{thiemannrev}
Thiemann T 2003 Lectures on loop quantum gravity \textit{Lect. Notes
Phys.} \textbf{631} 41-135 (\textit{Preprint:}
\texttt{gr-qc/0210094}

\bibitem{rovellibook}
Rovelli C 2004 \textit{Quantum Gravity} (Cambridge: Cambridge UP)

\bibitem{smolinrev}
Smolin L 2004 An invitation to loop quantum gravity
\textit{Preprint:} \texttt{hep-th/0408048}
%
% There is no journal reference for this in the arxiv entry.
% On the arXiv, it says
% ``submitted to Reviews in Modern Physics''; I checked the Review in
% in modern physics website, but I couldn't find it.
%
% I found a conference proceedings reference in a Ph.D. thesis
% on the web, and at least checked the pub. details for the conference
% proceedings:
%
% in \textit{Quantum Theory and Symmetries: Proceedings of the 3rd
% International Symposium (Cincinnati, 10-14 September 2003)},
% ed. Argyres P C et al. (World Scientific: Singapore, 2004), pp. 655-682
%
% However, I found this reference only in one obscure place, and
% I asked Lee if this is a correct reference for his review on the arXiv,
% and he said he doesn't think so, but he would try to check.  Therefore,
% I will not use this reference, but only the arXiv one.
%

\bibitem{einstein1956}
Einstein A 1956 \textit{The Meaning of Relativity} 5th ed.
(Princeton, NJ: Princeton UP), pp 55-56

\bibitem{barbero}
Barbero F 1995 Real Ashtekar variables for Lorentzian signature
space-times \textit{Phys. Rev.} D\textbf{51} 5507-5510
%
% I checked Abhay and Jurek's review, and they also
% cite Barbero as ``Barbero F''. (If you look at the name
% as given on the paper, it's confusing.)
%

\bibitem{rs_loops}
Rovelli C and Smolin L 1988 Knot theory and quantum gravity
\textit{Phys. Rev. Lett.} \textbf{61} 1155-1158 \\
Rovelli C and Smolin L 1990 Loop representation for quantum
general relativity \textit{Nucl. Phys.} B331 80-152

\bibitem{ai1992}
Ashtekar A and Isham C J 1992 Representations of the holonomy
algebras of gravity and non-Abelian gauge theories \textit{Class.
Quantum Grav.} \textbf{9} 1433-1468

\bibitem{al_repth}
Ashtekar A and Lewandowski J 1993
Representation theory of analytic holonomy C* algebras,
in \textit{Knots and Quantum Gravity}, ed. J Baez
(Oxford: Oxford UP)

\bibitem{al_diffgeom}
Ashtekar A and Lewandowski J 1995
Differential geometry on the space of connections via
graphs and projective limits
\textit{J. Geom. Phys.} \textbf{17} 191-230

\bibitem{al_area}
Ashtekar A and Lewandowski J 1996 Quantum theory of geometry I: Area
operators \textit{Class. Quantum Grav.} \textbf{14} A55-A82

\bibitem{lqcorig}
Bojowald M 2002 Isotropic loop quantum cosmology \textit{Class.
Quantum Grav.} \textbf{19} 2717-2742

\bibitem{abl2003}
Ashtekar A, Bojowald M and Lewandowski J 2003 Mathematical structure
of loop quantum cosmology \textit{Adv. Theor. Math. Phys.}
\textbf{7} 233-268 (\textit{Preprint:} \texttt{gr-qc/0304074})

\bibitem{ashtekar_cosm}
Ashtekar A 2007 An introduction to loop quantum gravity
through cosmology \textit{Preprint:} \texttt{gr-qc/0702030}

\bibitem{rudin1962}
Rudin W 1962 \textit{Fourier Analysis on Groups} (New York:
Interscience)

\bibitem{GCSI_II}
Thiemann T 2001 Gauge field theory coherent states (GCS): I. General
properties \textit{Class. Quantum Grav.}
\textbf{18} 2025-2064 \\
Thiemann T and Winkler O 2001 Gauge field theory coherent states
(GCS): II. Peakedness properties \textit{Class. Quantum Grav.}
\textbf{18} 2561-2636
%
% I checked again, and indeed Thomas's paper
% ``complexifier coherent states for QGR'' (below) refers
% to this series of papers as the papers generalizing the
% complexifier method to more general theories
% such as LQG. In the ``compl. coh. states. ...''
% paper, he simply says he is reviewing the material
% in the above paper, presenting it in a hopefully
% clearer way, and relating it to some other issues.
%

\bibitem{thiemann2006}
Thiemann T 2006 Complexifier coherent states for quantum general
relativity \textit{Class. Quantum Grav.} \textbf{23} 2063-2118
%
% There was also a paper published between the above two
% (in addition to the other GCS papers),
%
% Sahlmann H, Thiemann T, and Winkler O,
% ``Coherent States for Canonical Quantum General Relativity and
% the Infinite Tensor Product Extension.''
%
% I considered citing it for completeness, but some of the ideas in
% it are quite different from those used anywhere else, and it doesn't
% have any original results that are used in this paper.
% Furthermore, the above 2006 paper by Thomas doesn't even cite it,
% suggesting it is a direction that was abandoned (or which may even
% be wrong).  Thus I have not cited it, as it might only confuse things.
%

\bibitem{thiemann1996}
Thiemann T 1996 Reality conditions inducing transforms for quantum
gauge field theory and quantum gravity \textit{Class. Quantum Grav.}
\textbf{13} 1383-1404

\bibitem{bargmann}
Bargmann V 1961 \textit{Commun. Pure Appl. Math.} \textbf{14}
187-214
%
% \quad \textit{See also the topical review}\\
% Vourdas A 2006 Analytic representations in quantum mechanics
% \textit{J. Phys. A: Math. Gen.} \textbf{39} R65-R141

\bibitem{ashtekar1991}
Ashtekar A 1991 \textit{Lectures on Non-Perturbative Canonical Gravity}
(Singapore: World Scientific)

\bibitem{lqcmaster}
Bahr B and Thiemann T 2006 Approximating the physical inner product
of loop quantum cosmology \texttt{gr-qc/0607075}

% Am not using:
% \bibitem{bojspherical}
% Bojowald M 2004 Spherically symmetric quantum geometry: States and
% basic operators \textit{Class. Quantum Grav.} \textbf{21} 3733-3753
%

\bibitem{bf2007}
Brunnemann J and Fleischhack C 2007 On the configuration spaces
of homogeneous loop quantum cosmology and loop quantum gravity
\textit{Preprint:} \texttt{arXiv:0709.1621}

\bibitem{bt2006}
Brunnemann J and Thiemann T 2006 On (cosmological) singularity
avoidance in loop quantum gravity \textit{Class. Quantum Grav.}
\textbf{23} 1395-1427

\bibitem{thiemann1998}
Thiemann T 1998 Quantum spin dynamics (QSD) \textit{Class. Quantum
Grav.} \textbf{15} 839-873

%
% Note: for a minute, I thought I might cite Mikhail Kagan's
% article ``Phenomenological implications of an alternative
% Hamiltonian constraint for quantum cosmology'' as giving
% an ``alternative'' quantum Hamiltonian constraint, but in
% fact Mikhail doesn't talk about an alternative *quantum*
% Hamiltonian constraint, but rather stays at the level
% of effective equations.
%

\bibitem{bojinhom}
Bojowald M 2006 Loop quantum cosmology and inhomogeneities
\textit{Gen. Rel. Grav.} \textbf{38} 1771-1795

\bibitem{shadows}
Ashtekar A, Fairhurst S, and Willis J L 2003 Quantum gravity, shadow
states, and quantum mechanics \textit{Class. Quantum Grav.}
\textbf{20} 1031-1062

\bibitem{master}
Thiemann T 2006 Quantum spin dynamics VIII. The Master Constraint
\textit{Class. Quantum Grav.} \textbf{23} 2249-2266

\bibitem{almmt1995}
Ashtekar A, Lewandowski J, Marolf D, Mour\~{a}o J, Thiemann T
1995 Quantization of diffeomorphism invariant theories of connections
with local degrees of freedom
\textit{J. Math. Phys.} \textbf{36} 6456-6493
%
% Is the original paper discussing the refined algebraic
% quantization scheme and group averaging.
%
% (However Don Marolf
% a subsequent paper notes its equivalent to a procedure
% invented independently by a mathematician a couple years
% prior called ``Reiffel induction'' -- but the language
% used is completely different and hypermathematical and
% would thoroughly scare the reader. (Almost nobody in LQG cites
% the paper on Reiffel induction.) The reader would
% have no idea why I cited it, so I would have to explain,
% but its being cited for too minor a point to explain.
% More specifically, the purpose of the citation is
% just so the reader knows where to go to learn more about it;
% no result is being quoted.)
%


\bibitem{improved_dyn}
Ashtekar A, Pawlowski T and Singh P 2006
Quantum nature of the Big Bang: Improved dynamics
\textit{Phys. Rev.} D\textbf{74} 084003

\end{thebibliography}
\end{document}